\documentclass[a4paper,fleqn,usenatbib]{mnras}

\usepackage{newtxtext,newtxmath}

\usepackage[T1]{fontenc}
\usepackage{ae,aecompl}

\usepackage{graphicx}	
\usepackage{amsmath}	
 \usepackage{amssymb}	
\usepackage{color}
\usepackage{ulem}

\title[Appearance of a merging BBH close to a Kerr SMBH]{The appearance of a merging binary black hole very close to a spinning supermassive black hole}

\author[X. Zhang, X. Chen]{Xiaoyue Zhang$^{1}$, Xian Chen$^{1,2}\thanks{E-mail: \href{mail to: xian.chen@pku.edu.cn}{xian.chen@pku.edu.cn}} $
\\
$^{1}$Astronomy Department, School of Physics, Peking University, 100871 Beijing, China
\\
$^{2}$Kavli Institute for Astronomy and Astrophysics at Peking University, 100871 Beijing, China
}

\date{Accepted XXX. Received YYY; in original form ZZZ}

\pubyear{2023}

\begin{document}
\label{firstpage}
\pagerange{\pageref{firstpage}--\pageref{lastpage}}
\maketitle

\begin{abstract}
The mass and distance of a binary black hole (BBH) are fundamental parameters
to measure in gravitational-wave (GW) astronomy. It is well-known that the
measurement is affected by cosmological redshift, and recent works also showed
that Doppler and gravitational redshifts could further affect the result if the
BBH coalesces close to a supermassive black hole (SMBH).  Here we consider the
additional lensing effect induced by the nearby SMBH on the measurement.   We
compute the null geodesics originating within $10$ gravitational radii of a
Kerr SMBH to determine the redshift and magnification of the GWs emitted by the
BBH.  We find a positive correlation between redshift and demagnification,
which results in a positive correlation between the mass and distance of the
BBH in the detector frame.  More importantly, we find a higher probability for
the signal to appear redshifted and demagnified to a distant observer, rather
than blueshifted and magnified.  Based on these results, we show that a binary
at a cosmological redshift of $z_{\rm cos}=(10^{-2}-10^{-1})$ and composed of
BHs of $(10-20)M_\odot$ could masquerade as a BBH at a redshift of $z_{\rm
cos}\sim1$ and containing BHs as large as $(44-110)M_\odot$.  In the case of
extreme demagnification, we also find that the same BBH could appear to be at
$z_{\rm cos}>10$ and contain subsolar-mass BHs.  Such an effect, if not accounted for, 
could bias our understanding of the origin of the BHs detected via GWs. 
\end{abstract}




\section{Introduction}\label{sec:intro}

The majority of the black holes (BHs) detected by the Laser Interferometer
Gravitational-wave Observatory (LIGO) and the Virgo detectors \citep{GWTC3} are
several times more massive than those BHs found previously in X-ray binaries
\citep{mcclintock14,2016A&A...587A..61C}.  The discrepancy raises an
interesting question about the formation channel of the LIGO/Virgo binary BHs
(BBHs).  Conventional models suggest that massive BBHs could form as a result
of the evolution of isolated binary massive stars or the dynamical interactions
between the BHs in star clusters \citep[see][for
summaries]{2016ApJ...818L..22A,2019ApJ...882L..24A,2020ApJ...900L..13A}.

Another place where the LIGO/Virgo massive BBHs could form is the vicinity of a
supermassive black hole (SMBH, $10^6-10^{10}M_\odot$). First, massive objects
such as BHs could accumulate here due to a dynamical process called ``mass
segregation'' \citep{oleary_kocsis_2009}. It is also difficult for BHs to leave
the system because of the large escape velocity induced by the SMBH
\citep{MillerLauburg09}. Second, by interacting with the SMBH
\citep{antonini_perets_2012} or the other massive compact objects trapped in
the galactic nucleus \citep{2018MNRAS.474.5672L,2019ApJ...876..122Y}, a BBH
could more effectively merge.  Third, if the SMBH is surround by a gaseous
disk, as would be the case in an active galactic nucleus (AGN), BBHs could form
and merge even more efficiently by interacting with the gas
\citep[e.g.][]{baruteau11,mckernan_ford_2012,bartos_kocsis_2017,stone_metzger_2017}.
According to the estimations in the literature, the merger rate of the BBHs in
galactic nuclei is comparable to the rate inferred from the BBHs detected by
LIGO/Virgo
\citep{fragione19,tagawa_haiman_2019,ArcaSedda20,grobner20,ford21mckernan,gayathri21,zhang21,samsing22}.

More recent studies have shown that about $(1-2)\%$ of the LIGO/Virgo BBHs
could come from a distance smaller than $10$ Schwarzschild radii from a SMBH
\citep{chen_han_2018,addison_gracia-linares_2019,peng21}.  The distance could
be as small as $2-3$ gravitational radii if the central SMBH is rapidly
spinning \citep{chen22gem}.  Although such binaries constitute only a small
fraction of the BBH population, the fact that LIGO/Virgo have so far detected
$\sim90$ BBHs \citep{ligo21rate} indicates that one or two of them may have
come from the vicinity of a SMBH.  

Interestingly, a BBH so close to a SMBH
could appear significantly more massive than it really is \citep{chen_li_2019}.
This effect is analogous to the famous ``mass-redshift degeneracy''
\citep{schutz_1986} but now the redshift is caused not only by the expansion of
the universe, but also the gravitational and Doppler redshifts induced by the
SMBH.  The additional redshifts effectively lower the observed frequency ($f$)
and chirp rate ($\dot{f}$) of a gravitational-wave (GW) signal.  Since $f$ and
$\dot{f}$ are used to infer the mass of a BBH in the standard procedure of GW
data analysis, the inferred mass could be significantly larger than the intrinsic mass
if an observer, unaware of the presence of the SMBH, neglects the effects induced
by the gravitational and Doppler redshifts \citep{chen21book}.
 
Identifying these fake massive BBHs close to SMBHs is particularly important
given the current ambiguity of the origin of the LIGO/Virgo sources.  It is
also very challenging because the signal normally lasts less than a second in
the LIGO/Virgo band. The short duration of the signals, in contrast to the
long orbital period (tens to thousands of seconds) of the BBH around the SMBH, 
makes it difficult to discern any signature caused by the
acceleration of the GW source around the SMBH
\citep[e.g.][]{bonvin_caprini_2017,inayoshi_tamanini_2017,meiron_kocsis_2017}.

It has been noticed that the BBHs which appear excessively more
massive due to gravitational and Doppler redshifts would also appear more
distant \citep{chen_li_2019}. The cause is our convention of using the
ratio between the apparent chirp mass and the apparent GW amplitude to infer a
``luminosity distance'' for the source \citep{schutz_1986,holz_hughes_2005}.
Now that the chirp mass is overestimated, so is the luminosity distance.
Therefore, there seems to be a positive correlation between the apparent mass and the
apparent distance of a BBH, which may be used to search for the BBHs
in the vicinities of SMBHs.

However, very close to a SMBH, two additional relativistic effects would arise
and complicate the matter. First, gravitational lensing by the central SMBH
could magnify or demagnify the GW amplitude, making the source appear closer or
further away than its real luminosity distance. The effect of magnification was
noticed soon after Weber claimed detection of GW signals coming from the
Galactic Centre \citep{weber70} and was originally proposed to explain the
apparent high power of the GW source \citep{campbell73,lawrence73,ohanian73}.
More recent studies showed that the same effect can also produce echoes
\citep{kocsis13,gondan21,yu21} and repeated bursts of GWs \citep{dorazio20}.
However, the lens equation used in these earlier works is derived under the
assumption that the source is far from the lens
\citep{ohanian73,virbhadra00,bozza04}. The approximation is no longer valid
when the GW source falls inside a radius of $10$ Schwarzschild radii from the
central SMBH \citep{bozza08}. Moreover, the probability and observational
consequence of demagnification have not been fully accounted for in the
previous works. 

The second relativistic effect, that can affect the appearance of a BBH close to
a SMBH but has not been properly considered in the current model, is the
rotation of the SMBH. It is worth mentioning that the appearance of a star close to a spinning
SMBH is an old subject in relativistic astrophysics
\citep{polnarev72,cunningham73,ohanian73,1977ApJ...212..541P}. These earlier
works have shown that the spin generally enhances the degree of magnification
(or demagnification) of the light emitted from the star. In fact, GWs around a SMBH behave
just like light if we restrict ourselves to the waves detectable by LIGO/Virgo. The
wavelength is minuscule relative to the size a SMBH, making it a good
approximation to treat the trajectories of the GWs as null geodesics
\citep{isaacson_1967}. In addition, the short duration of a LIGO/Virgo event
allows us to further simplify the problem by neglecting the variation of the
position and velocity of the BBH around the SMBH.

Therefore, in this paper we will use the model designed to visualize a light
source around a SMBH \citep[e.g.][]{gralla18,thompson19} to study the
GW radiation of a BBH very close ($<10$ Schwarzschild radii) to a spinning SMBH.
The paper is organized as follows. In Section~\ref{sec:method} we 
describe the method that we use to solve the problem. 
We first use analytical arguments to verify how the measured chirp mass and
distance would change if the BBH is close to a 
SMBH. Then we lay down the equations of motion for the null geodesics around a SMBH, and
we use them to track the frequency and 
amplitude of the GWs as they propagates around the SMBH. 
In Section~\ref{sec:results} we show the resulting redshift and magnification 
of the GWs when they reach an observer in an arbitrary direction relative to the
source. In Section~\ref{sec:app} we show how the mass and distance will appear to observers. We give a brief summary of our result in Section~\ref{sec:dis} and discussion
their implication regarding the origin of the massive BBHs detected by LIGO/Virgo, as well 
as the detection of primordial BHs by future ground-based detectors.

\section{Methodology}  \label{sec:method}

\subsection{Measuring the mass and distance of a BBH}\label{sec:massdis}

We consider a BBH consisting of two stellar-mass BHs of masses $m_1$ and $m_2$,
where $m_1\ge m_2$. We assume that the binary is moving on a circular orbit in
the equatorial plane of a SMBH whose mass is $M$ and spin parameter is $a$
($0\le a<1$). Such a system could form in an AGN
\citep[e.g.][]{mckernan_ford_2012,peng21}. The distance between the BBH and the
SMBH is assumed to be small, so that the GWs emitted from the BBHs will be
affected not only by the normal cosmological redshift $z_{\mathrm{cos}}$, but
also a gravitational redshift $z_{\mathrm{gra}}$ and a Doppler redshift
$z_{\mathrm{dop}}$.

Take the inspiral part of the waveform for example.
Because of the redshift, the GW signal, when detected, will have a frequency of
$f_o=f(1+z)^{-1}$ and a chirp rate (time derivative of the frequency) of
$\dot{f}_o=\dot{f}(1+z)^{-2}$, where
$1+z=(1+z_{\mathrm{cos}})(1+z_{\mathrm{gra}})(1+z_{\mathrm{dop}})$ is the total
redshift, and $f$ and $\dot{f}$ are the corresponding values in the rest frame
of the BBH. From these two observables, one can infer a chirp mass of
\begin{equation} \label{1} 
\mathcal{M}_o=\frac{c^3}{G}\left(\frac{5f_o^{-11/3}\dot{f}_o}{96\pi^{8/3}}\right)^{3/5},
\end{equation} 
which uniquely determines how the GW frequency increases with time in the
detector frame. Here $G$ is the gravitational constant and $c$ the speed of
light.  It can be shown that $\mathcal{M}_o=\mathcal{M}(1+z)$, where
$\mathcal{M}=(m_1m_2)^{3/5}(m_1+m_2)^{-1/5}$ is the intrinsic chirp mass of the
BBH \citep[see][for details]{chen21book}. This result indicates that there is a
degeneracy between the mass and the redshift of the BBH. Without knowing the
redshift {\it a priori}, it is difficult to derive the intrinsic mass of the
BBH.

Besides $f_o$ and $\dot{f}_o$, there is a third observable which is the
amplitude of the GW, $h_o$. In the conventional scenario where both
$z_{\mathrm{gra}}$ and $z_{\mathrm{dop}}$ are negligible, the GW amplitude can
be used to infer the luminosity distance $d_L$ of the source, i.e., 
\begin{equation}
	d_L=\frac{4G}{c^2}\frac{\mathcal{M}(1+z_{\rm cos})}{h}\left(\frac{G}{c^3}\pi f\mathcal{M}\right)^{2/3},
	\label{eq:d_L}
\end{equation}
where we have used a different symbol $h$ to refer to the GW amplitude without
the presence of a SMBH. Here we have also neglected the dependence of $h$ on
the inclination of the orbital plane of the BBH, since we assume that future GW
observations could determine the inclination by measuring the two polarizations
of the GW signal \citep{sathyaprakash_schutz_2009}. The luminosity distance can
be used to further infer the cosmological redshift $z_{\mathrm{cos}}$ if a
cosmological model is provided. Using the $z_{\mathrm{cos}}$ inferred from the
luminosity distance, one can eventually break the mass-redshift degeneracy and
derive the intrinsic chirp mass $\mathcal{M}$ of the BBH.

The apparent distance would become different if the BBH resides in the vicinity
of a SMBH.
Now the amplitude of the observed GW signal, $h_o$,
could be greater or smaller than $h$, depending on whether the signal gets magnified or
demagnified by the SMBH. As a result, the distance of the source appears to be
\begin{equation}
d_o=\frac{4G}{c^2}\frac{\mathcal{M}_o}{h_o}\left(\frac{G}{c^3}\pi f_o\mathcal{M}_o\right)^{2/3},
	\label{eq:d_o}
\end{equation}
which is different from $d_L$.  Noticing that
$1+z=1+z_{\mathrm{cos}}$  in Equation~(\ref{eq:d_L}) but
$1+z=(1+z_{\mathrm{cos}})(1+z_{\mathrm{gra}})(1+z_{\mathrm{dop}})$
in Equation~(\ref{eq:d_o}), we find that
\begin{equation}
	\frac{d_o}{d_L}=(1+z_{\mathrm{dop}})(1+z_{\mathrm{gra}})\frac {h}{h_o}.\label{eq:dodL}
\end{equation}
Depending on the value on the right-hand side of the last equation, the source
could appear either closer or more distant than its real luminosity distance ($d_L$).
We note that the factor of $h/h_o$ has been missing in the previous calculations of $d_o$ 
\citep{chen_li_2019,chen21book,vijaykumar22,torres-orjuela22hubble} since these
earlier works did not consider the (de)magnification of GWs by the SMBH. 

An observer, unaware of the presence of the SMBH, has a tendency of using $d_o$
to infer the cosmological redshift of the source. The result, which we denote
by $z_{d_o}$, deviates from the real cosmological redshift $z_{\mathrm {cos}}$.
Consequently, the chirp mass inferred from $z_{d_o}$, i.e.,
\begin{equation}
    \mathcal{M'}=\frac{\mathcal{M}_o}{1+z_{d_o}},
    \label{eq:Mrelation}
\end{equation}
also differs from the real chirp mass $\mathcal{M}$ of the source. In fact,
\begin{equation}
    \frac{\mathcal{M'}}{\mathcal{M}}=(1+z_{\mathrm{dop}})(1+z_{\mathrm{gra}})
	\left(\frac{1+z_{\mathrm{cos}}}{1+z_{d_o}}\right).\label{eq:Mprime}
\end{equation}
The factor $(1+z_{\mathrm{cos}})/(1+z_{d_o})$, which corrects for the effect of
(de)magnification, has not been considered in the previous works either.

\subsection{Equations of motion for GWs}

Equations~(\ref{eq:dodL}) and (\ref{eq:Mprime}) indicate that the mass and the
distance of a BBH will appear very different if the binary resides close to a
SMBH. To determine the values of $z_{\mathrm{gra}}$, $z_{\mathrm{dop}}$ and $h/h_o$
in these equations, we must first understand how GWs propagate around a SMBH.
To facilitate the calculation, we split the space-time into two regions. 

The first region  is centred on the SMBH with a relatively small scale compared
to the size of the universe.  Inside this region, the space-time is dominated
by the Kerr metric which, in the Boyer-Lindquist coordinate, can be written as
\begin{equation}
    \begin{aligned}
    \mathrm{d}s^2=&-\Big(1-\frac{2Mr}{\Sigma}\Big)\mathrm{d}t^2-\frac{4Mar\sin^2\theta}{\Sigma}\mathrm{d}t\mathrm{d}\phi+\frac{\Sigma}{\Delta}\mathrm{d}r^2\\
    &+\Sigma\mathrm{d}\theta^2+\sin^2\theta\Big(r^2+a^2+\frac{2Ma^2r\sin^2\theta}{\Sigma}\Big)\mathrm{d}\phi^2,
    \end{aligned}
\end{equation}
where
\begin{equation}
    \Sigma\equiv r^2+a^2\cos^2\theta,\qquad\Delta\equiv r^2-2Mr+a^2.
\end{equation}
Here and in the following analysis, we use the geometrized units where $G=c=1$. 

Such a metric induces the Doppler effect, the gravitational redshift and the
(de)magnification of GWs.  Since we are interested in the GWs in the LIGO/Virgo
band, the wavelength ($10^3-10^5~\mathrm{km}$) is much shorter than the
curvature radius of the SMBH ($M\ga10^6M_\odot$). Therefore, we are safely in
the geometric-optics limit and can approximate the trajectories of the GWs by
null geodesics \citep{isaacson_1967}.

The geodesic in Kerr metric allows three constants of motion, namely, 
energy $E=-p_{t}$, angular momentum $L=p_{\phi}$ and the Carter constant
\begin{eqnarray}
	Q&=&p_{\theta}^2+p_{\phi}^2\cot^2\theta-a^2(p_t^2-\mu^2)\cos^2\theta\\
	&=&p_{\theta}^2+L^2\cot^2\theta-a^2(E^2-\mu^2)\cos^2\theta
\end{eqnarray}
\citep{1968PhRv..174.1559C,2009IJMPD..18..429R,2020PhRvD.101d4032G}.
We note that for null geodesics, $\mu^2=-k_{\mu}k^{\mu}=0$ where
$k^\mu$ is the wave vector.
Using these three constants, we can write the equations of motion of the geodesic
as
\begin{subequations}
\begin{align}
\dot{r}&=\frac{\Delta}{\Sigma}p_r,\\
\dot{p}_r&=-\left(\frac{\Delta}{2\Sigma}\right)'p_r^2-\left(\frac 1{2\Sigma}\right)'p_{\theta}^2+\left(\frac{R+\Delta\Theta}{2\Delta\Sigma}\right)',\\
\dot{\theta}&=\frac 1{\Sigma}p_{\theta},\\
\dot{p}_{\theta}&=-\left(\frac{\Delta}{2\Sigma}\right)^{\theta}-\left(\frac 1{2\Sigma}\right)^{\theta}p_\theta^2+\left(\frac{R+\Delta\Theta}{2\Delta\Sigma}\right)^{\theta},\\
\dot{t}&=\frac 1{2\Delta\Sigma}\frac{\partial}{\partial E}(R+\Delta\Theta),\\
\dot{p}_t&=0,\\
\dot{\phi}&=-\frac1 {2\Delta\Sigma}\frac{\partial}{\partial L}(R+\Delta\Sigma),\\
\dot{p}_{\phi}&=0,
\end{align}\label{eq:eom}
\end{subequations}
where the dots are derivatives with respect to the affine parameter $\lambda$, 
the primes are derivatives with respect to $r$, superscripts $\theta$ are derivatives with respect to $\theta$, and the functions $R$ and $\Theta$
are defined as
\begin{subequations}
     \begin{align}
     R(r)&\equiv(E(r^2+a^2)-aL)^2-\Delta(r)[\mu^2r^2+Q+(L-aE)^2],\\
     \Theta(\theta)&\equiv Q+a^2(E^2-\mu^2)\cos^2\theta-L^2\cot^2\theta.
     \end{align}
\end{subequations}

These equations of motion were derived by \citet{2008PhRvD..77j3005L} to avoid
the numerical problems related to turning points and square roots.  Although
they were derived for timelike geodesics, they also apply to null geodesics as
long as the constants of motion are properly chosen. Using these equations, we
compute the null geodesics to a distance of $5\times10^6M$, which is big
compared to the size of the SMBH but small relative to the size of the
universe.

The second region is at $r>5\times10^6M$. We assume that the space-time here is
dominated by the Friedmann-Lema\^itre-Robertson-Walker (FLRW) metric. This region
is causing the cosmological redshift, as well as a decay of the GW amplitude with
the luminosity distance according to Equation~(\ref{eq:d_L}).

\subsection{Calculating the redshift}\label{sec:red}

The redshift of the GW seen by a distant observer is calculated in two steps.
In the first step, we use the equations of motion specified in the last
subsection to propagate the 4-momentum $p_\mu$ to the edge of the first region,
i.e., $r=5\times10^6M$. This $p_\mu$ allows us to calculate the Doppler and
gravitational redshifts, as we will elaborate below.  In the second step, we
add the cosmological redshift to account for the effect induced by the
expansion of the universe in the second region.

Now we describe the details of calculating the Doppler and gravitational
redshifts in the first step. During the propagation of the GW, its energy seen
by a local observer is $-(p_{\mu}U^{\mu})_{\mathrm{obsv}}$, where $U^{\mu}$ is
the 4-velocity of the local observer.  Given the Kerr metric, it is convenient
for our later calculations if the local observer is chosen to coincide with a
special frame called ``the locally non-rotating frame''
\citep[LNRF,][]{1970ApJ...162...71B,1972ApJ...178..347B,1973grav.book.....M}.
Using the above expression for the GW energy, we can write the combined
redshift as
\begin{equation}
	(1+z_{\rm dop})(1+z_{\rm gra})=\frac{(p_{\mu}U^{\mu})_{\mathrm{emit}}}{(p_{\mu}U^{\mu})_{\mathrm{obsv}}},
\end{equation}
where the term $(p_{\mu}U^{\mu})_{\mathrm{emit}}$ comes from the energy of the GW
in the rest frame of the source.

The numerator and denominator in the last equation are determined as follows. 
For the source (emitter), we assume that it moves on a circular orbit around the
SMBH because of the reason given at the beginning of Section~\ref{sec:massdis}.
Therefore, we can write 
\begin{equation}
(p_{\mu}U^{\mu})_{\mathrm{emit}}/E=(U^t-bU^{\phi})_{\mathrm{emit}}=U^t(1-b\Omega_s)_{\mathrm{emit}},
	\label{eq:Eemit}
\end{equation}
where $U^t$ and $U^{\phi}$ are, respectively, the time and azimuthal components
of the 4-velocity of the source, $\Omega_s\equiv U^{\phi}/U^t$ is the angular
velocity of the source, and $b\equiv {L}/{E}$.
To calculate $\Omega_s$ and $U^t$, we use the equations 
\begin{equation}
    \begin{aligned}
    \Omega_s&=\pm\frac{M^{1/2}}{r_s^{3/2}\pm aM^{1/2}},\\
    U^t&=\gamma\sqrt{\frac{\Xi}{\Delta\Sigma}}
    \end{aligned}
\end{equation}
\citep{1972ApJ...178..347B,gralla18}, where
$\gamma=1/\sqrt{1-v_s^2}$ is the Lorentz factor of the source, $r_s$ is the 
distance between the source and the SMBH, $v_s=\Xi(\Omega_s-\omega)/(\Sigma\sqrt{\Delta})$ is the 3-velocity of the source relative to the LNRF, $\omega=2aMr/\Xi$ is the angular velocity of the LNRF, and $\Xi=(r^2+a^2)^2-\Delta a^2\sin^2\theta$.
 For the denominator (observer), we can write
\begin{equation}
	(p_{\mu}U^{\mu})_{\mathrm{obsv}}/E=U^0(1-b\omega)_{\mathrm{obsv}},\label{eq:Eobsv}
\end{equation}
where $U^0=\sqrt{\Xi/(\Delta\Sigma)}$ is the time component of the 4-velocity of the
LNRF. Combining Equations~(\ref{eq:Eemit}) and (\ref{eq:Eobsv}), we find that
\begin{equation}
	(1+z_{\rm dop})(1+z_{\rm gra}) =\frac{U^t(1-b\Omega_s)_{\mathrm{emit}}}{U^0(1-b\omega)_{\mathrm{obsv}}}.\label{eq:zgradop}
\end{equation}

\begin{figure*}
    \centering
    \begin{tabular}{cc}
        \includegraphics[width=0.5\textwidth]{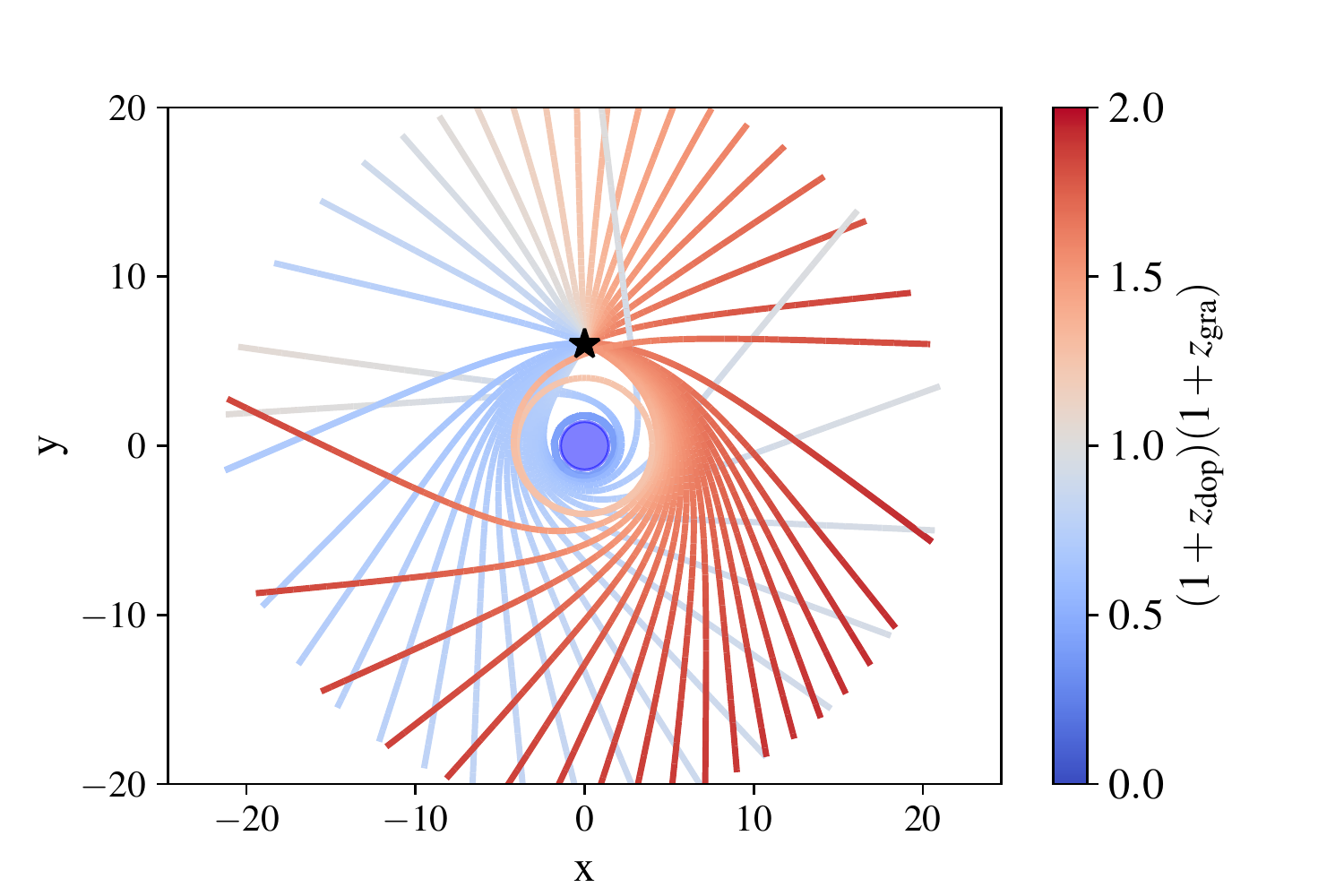}&
        \includegraphics[width=0.5\textwidth]{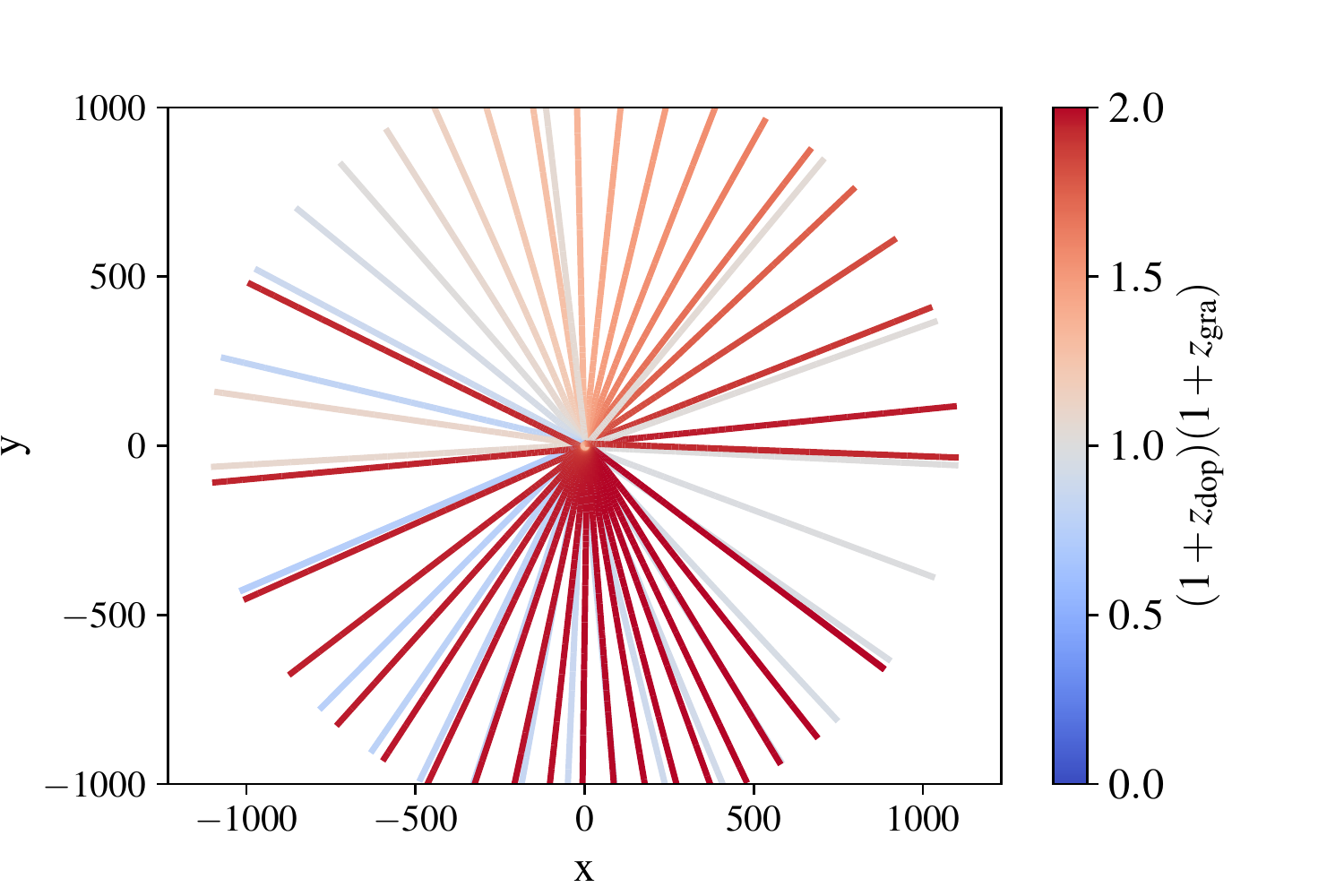}
    \end{tabular}    
	\caption{Trajectories of the GWs emitted by a BBH (black star) which is moving along a circular 
	orbit ($r_s=6M$) in the equatorial plane of a Kerr SMBH (a=0.9). The color shows the frequency
	shift due to the Doppler effect and gravitational redshift seen in the LNRF. A red color indicates a redshift and blue indicates
	a blueshift. The left panel shows the central part of the equatorial plane with a size of
	about $20M\times20M$. The right one shows the outer part of $10^3M\times10^3M$. The black star in the left panel shows the position of the binary source.
}
    \label{fig:localz}
\end{figure*}

For example, 
Figure~\ref{fig:localz} shows the null geodesics starting from a source in the equatorial plane
of a Kerr SMBH. Here we only show the waves propagating in the equatorial plane. From the left panel,
we can clearly see the bending of the null geodesics due to the lensing effect.

The color is calculated according to Equation~(\ref{eq:zgradop}) and shows how
the GW frequency shifts as the wave propagates towards infinity.  We find that
both redshift ($1+z>1$) and blueshift ($0<1+z<1$) can happen. The latter is
correlated with those null geodesics which are emitted close to the direction
of the orbital velocity of the source (see the left panel). These GWs are
initially blueshifted due to the Doppler effect.  

Moreover, we find that at large distances (see the right panel), the magnitude
of redshift is on average larger than the magnitude of blueshift.  The effect
is caused by the fact that gravitational redshift affects all geodesics.  Such
an asymmetry between red and blueshift could be partially responsible for the
observational result that there are more LIGO/Virgo BHs in the upper mass gap
than in the lower one, since the chirp mass in the detector frame is
$\mathcal{M}(1+z)$. 

Another interesting result is that at large distances (right panel) a
redshifted null geodesic could overlap with a blueshifted one. An observer in
such a special direction could detect two ``images'' with a time delay, as
previous works have pointed out \citep{kocsis13,gondan21,yu21}. However, our
result indicates that the two images will appear to have different masses in
the detector frame (also different distances, see Section~\ref{sec:res_mag}).
Therefore, they are more likely to be identified as physically unrelated
separate events by future observations.

In real observations, the observer is almost at the infinity relative to the
source.  Therefore, Equation~(\ref{eq:zgradop}) reduces to $(1+z_{\rm
dop})(1+z_{\rm gra}) \simeq U^t(1-b\Omega_s)_{\mathrm{emit}}$ because
$U^0\simeq1$ and $\omega\simeq0$ as $r$ becomes infinity.  In the following, we
use such a reduced equation to compute the Doppler and gravitational redshift
relative to a distant observer.  Notice that the equation only depends on the
constants of motion through the impact parameter $b$.  Therefore, by varying
$b$ in the physically allowed range, we can derive
the upper and lower limits of $(1+z_{\rm dop})(1+z_{\rm gra})$ without
integrating the geodesic equations. 

Finally, if we also know the cosmological redshift $z_{\rm cos}$ of the source,
we can calculate the total frequency shift of the GW with $1+z=(1+z_{\rm
cos})(1+z_{\rm dop})(1+z_{\rm gra})$.

\subsection{Calculating the magnification factor}\label{sec:mag}

To calculate the magnification factor $h_o/h$ which first appeared in
Equation~(\ref{eq:dodL}), we follow the method given in
\citet{campbell73} and consider a ray bundle coming out of the emitter
and propagating along null geodesics. Along each ray, the frequency $\nu$
of the GW, the energy flux ${\cal F}$, and the cross section $\mathrm{d}A$ of the ray
bundle satisfy the relationship
\begin{equation}
	\nu_e^{-2}{\cal F}_e \mathrm{d}A_e=\nu_o^{-2}{\cal F}_o \mathrm{d}A_o,\label{eq:law}
\end{equation}
which results from the area intensity law. Here we have used the subscript $e$
to denote the quantities in the rest frame of the emitter, and the subscript
$o$ in the frame of the observer (detector). In particular, we assume that both
${\cal F}_e$ and $\mathrm{d}A_e$ are evaluated at a distance from the BBH that
is much smaller than the curvature radius ($\sim M$) of the SMBH. Therefore, we can
use the equations derived in flat (Minkowski) space-time to calculate $h_e$. 
Substituting the expression in flat space-time,
\begin{equation}
	{\cal F}=\frac{4Gh^2\nu^2}{\pi c^3}\label{eq:flux},
\end{equation}
for the fluxes in Equation~(\ref{eq:law}), we find that
\begin{equation}
	h_e^2\mathrm{d}A_e=h_o^2\mathrm{d}A_o.\label{eq:newLaw}
\end{equation}

To relate $h_e$ to $h$, the amplitude of the GW if there is no SMBH, we notice
that the left-hand side of the above equation is evaluated in the rest frame of
the source.  Suppose the ray bundle in this frame span a solid angle of
$\mathrm{d}\Omega_e$ relative to the source, we can use the area intensity law
again when there is no SMBH and derive $h_e^2 d_e^2\mathrm{d}\Omega_e=h^2
d_C^2\mathrm{d}\Omega_e$, where $d_e$ is a small (proper) distance relative to
the source ($d_e\ll M$ in our model) and $d_C$ is the transverse comoving
distance between the BBH and the observer.

Now we can use the relationship between $h_e$ and $h$ to replace the $h_e$ in
Equation~(\ref{eq:newLaw}). Before doing that, we further notice that
$\mathrm{d}A_o\simeq d_C^2\mathrm{d}\Omega_o$, where $\mathrm{d}\Omega_o$ is
the solid angle of the ray bundle relative to the SMBH when viewed by a
distance observer. This approximation is appropriate because we have seen in
Figure~\ref{fig:localz} that the rays become straight lines when they propagate
to large distances from the SMBH. Combining the relationships above, we find that
Equation~(\ref{eq:law}) reduces to
\begin{equation}
    \frac{h_o}{h}\simeq\sqrt{\frac{\mathrm{d}\Omega_e}{\mathrm{d}\Omega_o}}.
    \label{eq:ho}
\end{equation}
According to this equation, the GWs will be magnified if the ray bundle becomes
more focused after being lensed by the SMBH.  We note that although frequency
does not explicitly appear in the above equation because $\nu$ is canceled in Equations (\ref{eq:law}) and (\ref{eq:flux}), the equation has properly
accounted for the frequency shift due to the Doppler effect, gravitational
redshift and the expansion of the universe.

Now the remaining task is to to calculate $\mathrm{d}\Omega_e$ and
$\mathrm{d}\Omega_o$ spanned by the ray bundle.  To express
$\mathrm{d}\Omega_o$, we use the Boyer-Lindquist coordinates $\theta$ and
$\phi$. Then we can write $\mathrm{d}\Omega_o=\sin\theta \mathrm{d}\theta
\mathrm{d}\phi$.  For $\mathrm{d}\Omega_e$, we define the usual spherical
coordinates $\theta_e-\phi_e$ in the rest frame of the source and write
$\mathrm{d}\Omega_e=\sin\theta_e \mathrm{d}\theta_e \mathrm{d}\phi_e$.  To
quantify the relationship between the angles $\theta$ and $\phi$ in the
SMBH's frame and the angles $\theta_e$ and $\phi_e$
in the source frame, we use the fact that along each ray two quantities, namely
the rescaled constants $b$
and $q=\sqrt{Q}/E$, are both invariant. 

In practice,  $\mathrm{d}\Omega_e$ and $\mathrm{d}\Omega_o$ are calculated
numerically as follows.  
Given a direction $(\theta_e,\phi_e)$ in the source frame, we first
calculate $b$ and $q$ of the ray using the method specified in \citet{gralla18} 
\footnote{We did not directly use the angles defined in \citet{gralla18}, but converted them to 
the usual spherical coordinates
$(\theta_e,\phi_e)$.}.  Then
using $b$ and $q$, we integrate the null geodesic in the Boyer-Lindquist
coordinates to a large distance ($r=5\times10^6M$ in our model) to derive
$\theta$ and $\phi$. Notice that in a fraction of the parameters space of $b-q$, 
the rays will fall into the central SMBH and cannot escape to infinity
\citep{2021PhRvD.103j4028I}. Using the first ray as a reference, we also
calculate two neighbouring rays emitted in slightly different directions 
in the source frame, i.e.,
$(\theta_e+\Delta\theta_e,\phi_e)$ and $(\theta_e,\phi_e+\Delta \phi_e)$
where $\Delta\phi_e=5\times10^{-3}$ and $ \Delta\theta_e=5\times10^{-4}$.
When these two rays reach a distance of $r=5\times10^6M$, their Boyer-Lindquist
coordinates are recorded as $(\theta_1,\phi_1)$ and $(\theta_2,\phi_2)$.
Since both $\Delta\theta_e$ and $\Delta\phi_e$ are small, we use
the equations
\begin{equation}
    \Delta\Omega_e
	=\Delta\phi_e\left|\cos(\theta_e+\Delta\theta_e)-\cos\theta_{e}\right|/2
\end{equation}
and
\begin{equation}
    \Delta\Omega_o=\left|(\phi_1-\phi)(\cos\theta_2-\cos\theta)-(\phi_2-\phi)(\cos\theta_1-\cos\theta)\right|/2
\end{equation}
to compute the solid angles, respectively, at the source (in the source frame) and at the observer (in the SMBH's frame).

\section{Results}\label{sec:results}

\subsection{Magnification as a function of viewing angle}\label{sec:res_mag}

We first study the magnification of GW.  We
choose $41553$ directions which are uniformly sampled in the rest frame of the
source. We then calculate the solid angles spanned by the ray bundles using the
method described in Section~\ref{sec:mag}. Using the solid angles in the source
frame and the detector frame, we derive the magnification factors along those
$41553$ directions.

\begin{figure}
	\includegraphics[width=0.45\textwidth]{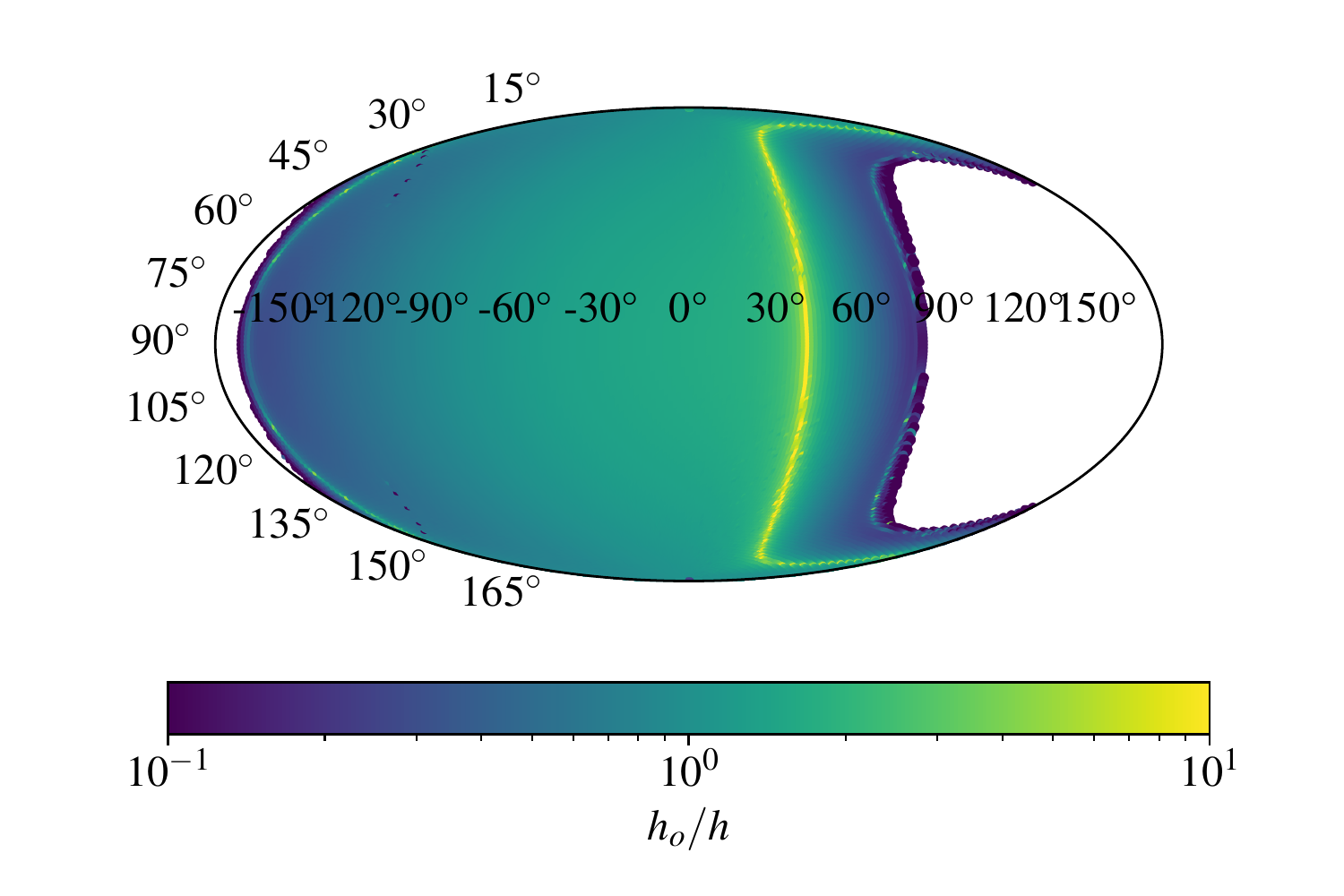}
	\caption{Dependence of the magnification factor $h_o/h$ on the direction 
	$(\theta_e,\phi_e)$ of the ray when viewed
	in the rest frame of the source. The parameters are $a=0.9$ and $r_s=3M$.
}\label{fig:mag}
\end{figure}

First, we stay in the source frame and study the magnification as a function of the 
direction of the ray. The result is shown in 
Figure~\ref{fig:mag}. In this
example, we set $a=0.9$ and $r_s=3M$.  The azimuthal angle $\phi_e=0$
corresponds to the direction of the orbital velocity of the source (BBH) and
the inclination angle $\theta_e=90^\circ$ 
corresponds to the equatorial plane of the Kerr SMBH.

We find the following interesting features in Figure~\ref{fig:mag}. (i) The
most obvious one is a large region with white color. This region corresponds to
the directions in which the null geodesics eventually fall into the central
SMBH \citep{2021PhRvD.103j4028I}.  (ii) The edge of the white region becomes
dark blue, indicating that the GWs in these directions are demagnified.  In
fact, many rays here can circle around the SMBH multiple
times before they can escape to infinity. During this process, the
magnification factor decays exponentially with the number of cycles the ray
goes around the SMBH \citep{kocsis13}.  (iii) Another prominent feature is a
ring with bright yellow color circling the white region. The yellow color
indicates large magnification. As we will show later, inside the ring
neighbouring null geodesics will intersect.  These behaviours resemble the
behaviours of the light rays around a caustic in optics
\citep{1992grle.book.....S,1996CQGra..13.1161H}. For this reason, we call the bright
yellow ring in Figure~\ref{fig:mag} ``caustic'' as well.

\begin{figure}
    \includegraphics[width=0.45\textwidth]{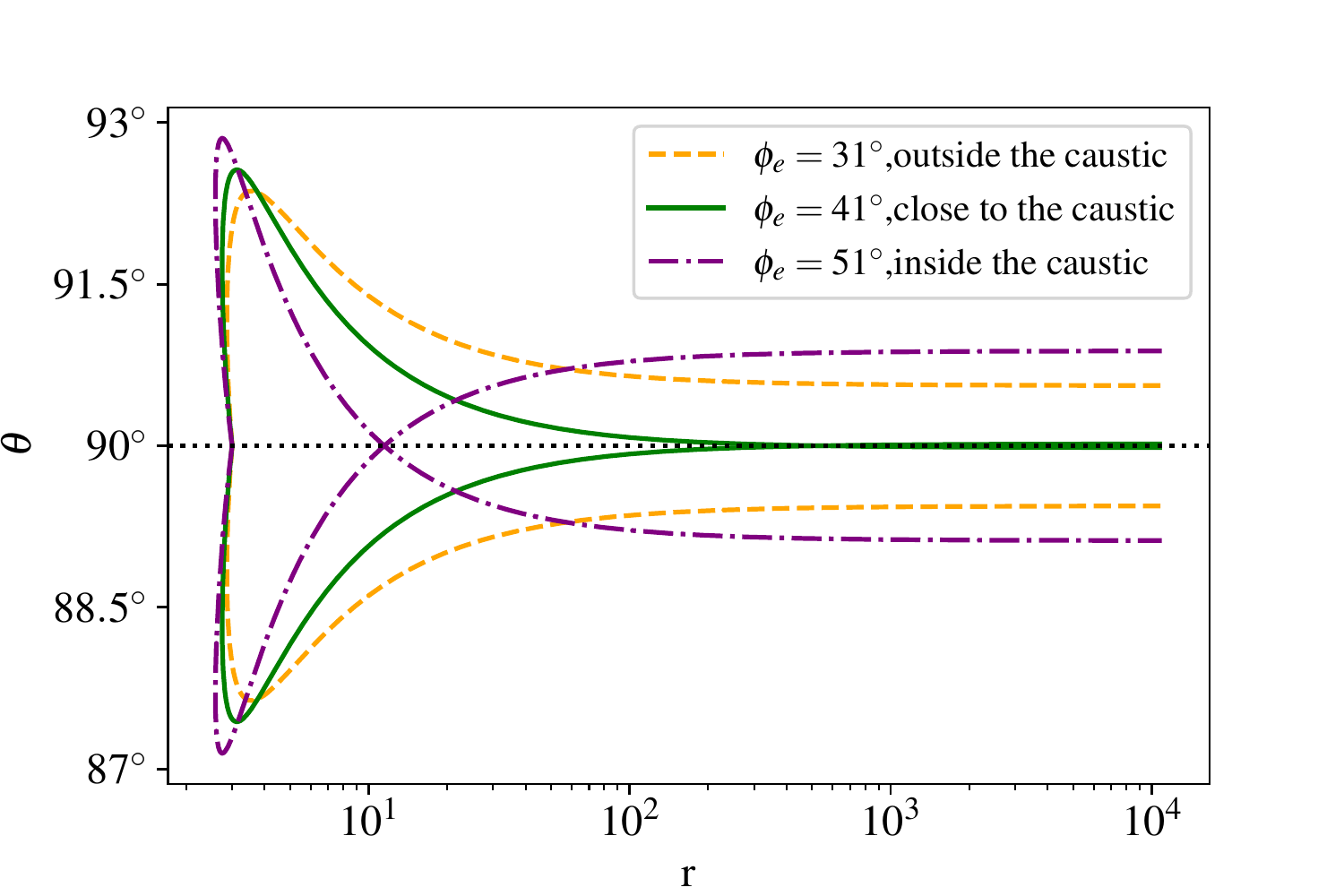}
	\caption{Evolution of the null geodesics close to the equatorial plane of the Kerr SMBH. 
	Different line styles correspond to different initial $\phi_e$ values in the rest frame
of the BBH.
}
    \label{caustic}
\end{figure}

To better understand the cause of the caustic, we show in
Figure~\ref{caustic} the evolution of the $\theta$ value of a null geodesic as
it propagates to larger $r$. For illustrative purposes, we restrict ourselves
to the rays close to the equatorial plane, i.e., we consider two different
values of $\theta_e$, $90^\circ\pm\Delta\theta_e$ where
$\Delta\theta_e\ll10^\circ$.  In addition, we choose three different $\phi_e$
directions (in the source frame), corresponding to the rays outside, close to
and inside the caustic. We find that the rays close to the caustic (green solid
lines) converge to nearly one point. That is why the magnification is high
close to the caustic. Moreover, the rays inside the caustic (purple dash-dotted
lines) intersect with the equatorial plane and end up in a hemisphere opposite
to the one from which the rays originate. 
 
Two additional features in Figure~\ref{fig:mag} are worth mentioning. (i) Between the
bright yellow ring and the edge of the white region is another
narrower yellow ring. We find that the geodesics starting from the region
between the two yellow rings will
intersect with their neighbour geodesics only in one direction,
either $\theta$ or $\phi$. However, the geodesics inside the inner yellow ring can
intersect with their neighbours in both directions.  (ii) We also see
several dark blue spots in the upper left and lower left directions. These rays
are close to spherical photon orbits relative to the SMBH. They appear
only when the source is inside the photon circular orbit in the retrograde direction.  They also
circle around the SMBH multiple times before escaping to
infinity, which results in their strong demagnification.

\begin{figure*}
\centering
\begin{tabular}{cc}
    \includegraphics[width=0.45\textwidth]{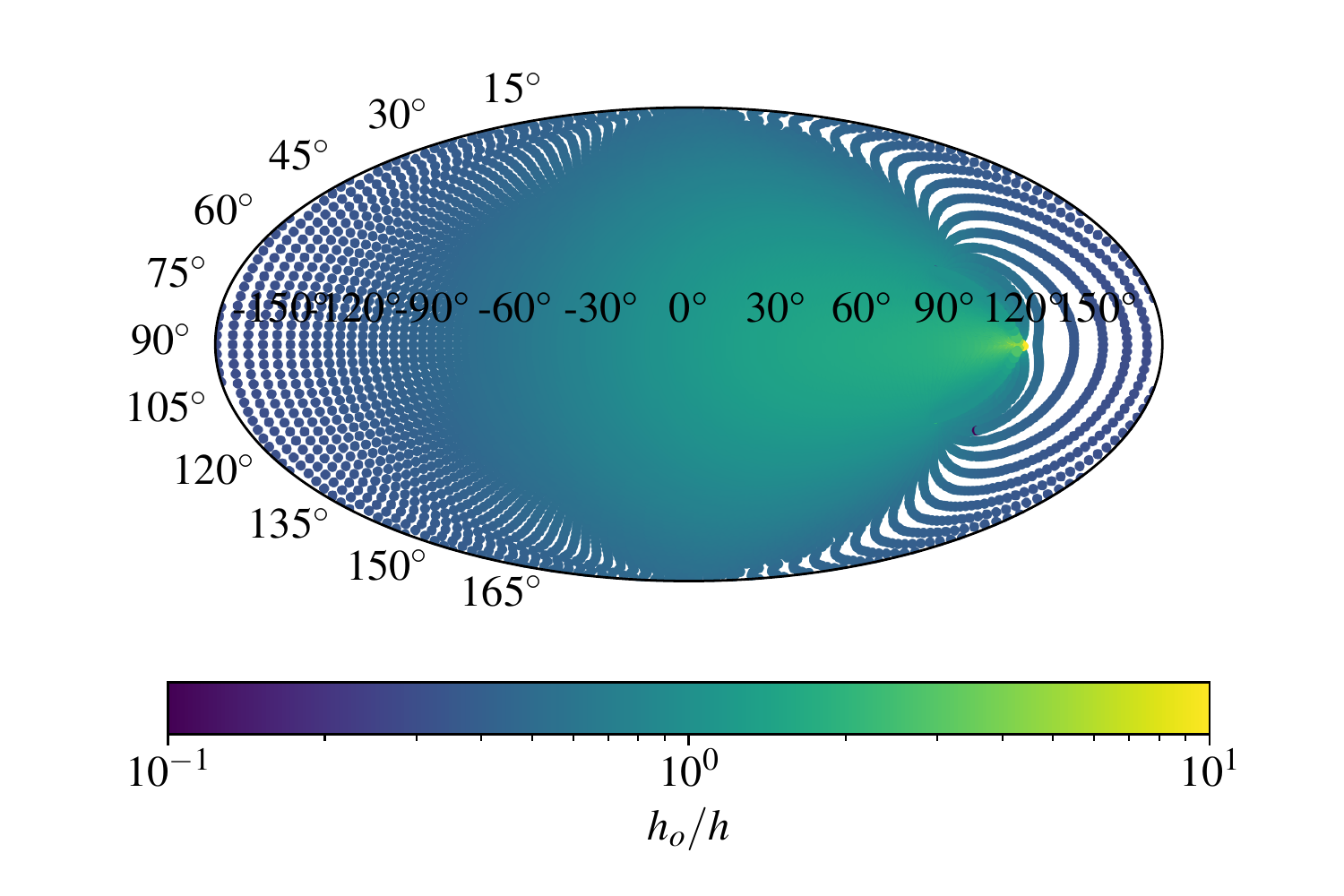}&
    \includegraphics[width=0.45\textwidth]{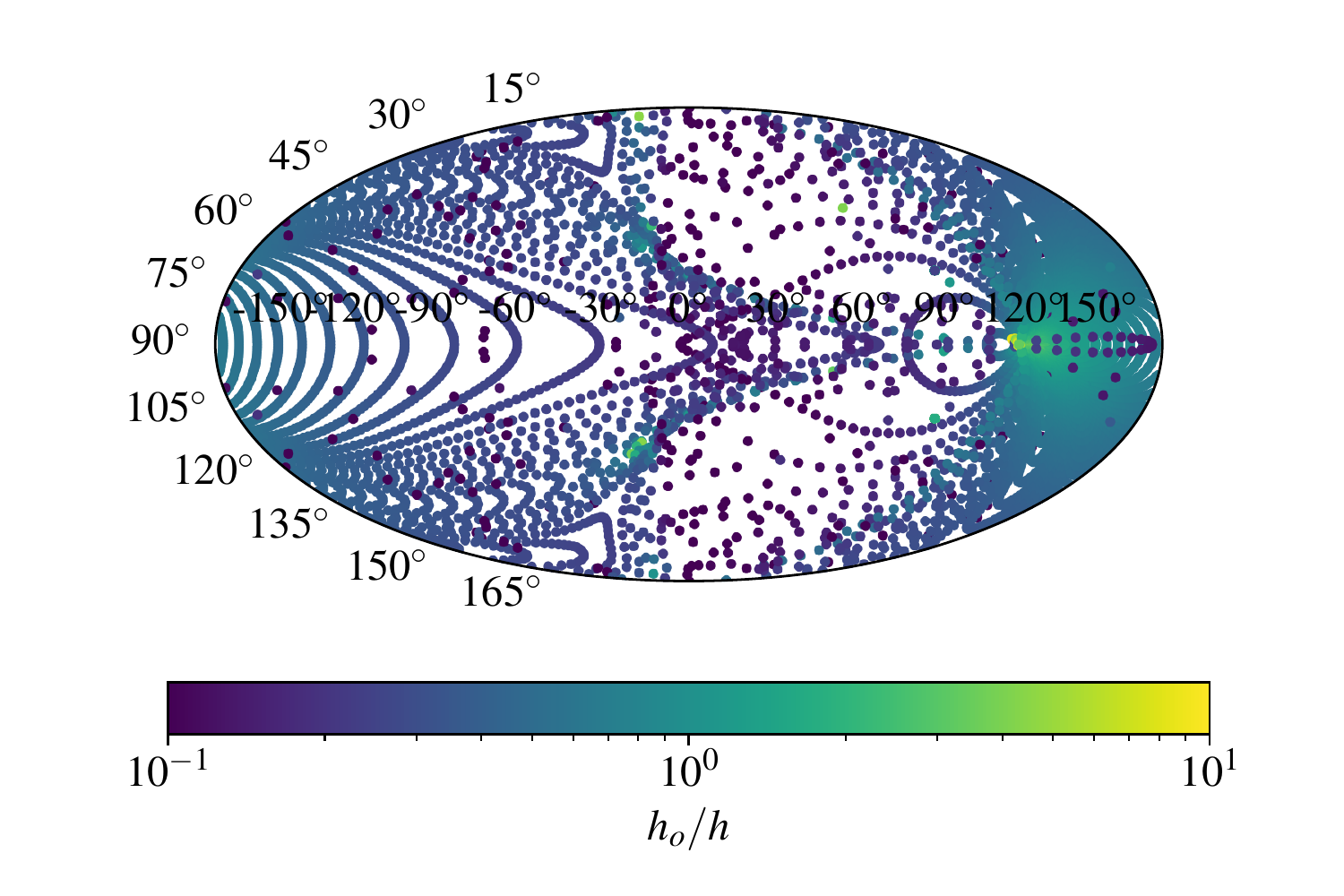}
\end{tabular}
	\caption{Magnification $h_o/h$, seen by distant observers, 
as a function of the Boyer-Lindquist coordinates $(\theta,\phi)$
centred on the Kerr SMBH.
	The parameters are the same as in Figure~\ref{fig:mag}. 
	Here the inclination angle $\theta=90^\circ$ corresponds to the 
	equatorial plane
and we have chosen the azimuthal angle $\phi=0$ to be the direction of the
orbital velocity of the BBH. In this coordinate system, an observer at $\theta=90^\circ$
and $\phi=90^\circ$ is on the opposite side of the SMBH when viewed form the BBH.
	. The left panel shows the rays
starting from the exterior of the bright yellow circle (caustic) in Figure~\ref{fig:mag}. The
right panels shows the rays originating inside the caustic.
}
    \label{hdis}
\end{figure*}

The above sky map of magnification is viewed in the source frame. When the rays arrive at a distant observer, however, the
angular dependence of the magnification will look different due to the bending
of the null geodesics around the SMBH. A natural coordinate system to
visualize the distribution is the Boyer-Lindquist coordinate.
Figure~\ref{hdis} shows the magnification as a function of the Boyer-Lindquist
coordinate $(\theta,\phi)$ of the SMBH.  The sky direction in this plot is
directly connected to the line-of-sight of a distant observer.  Notice that we
have chosen $\phi=0^\circ$  to be the direction of the orbital velocity of the 
BBH in the
Boyer-Lindquist coordinates. To see more clearly the role of the caustic, we
plot in the left panel only those rays originating from the exterior of the
caustic (out of the bright yellow ring in Figure~\ref{fig:mag}). The rays
originating inside the caustic are shown in the right panel. 

From the left panel we see a pattern that is very different from
Figure~\ref{fig:mag}.  We no longer see a large region with white color.
Instead, we find white gaps, but they are numerical, caused by the limited
number of rays that we have chosen.  This result indicates that the distant
observer can ``see'' the BBH from all angles.  Even if the observer and the BBH
are on opposite sides of the SMBH, one ray from the BBH can still go around the
SMBH, due to the space-time curvature, and reach the observer.

From the same panel, we also find that the region of strong magnification
($h_o/h\simeq10$) is concentrated in one direction, close to $\phi=120^\circ$
and $\theta=90^\circ$. Notice that there is a large offset between this
direction of maximum magnification and the velocity of the BBH ($\theta=0$,
$\phi=0$).  The solid angle of this region where magnification is prominent is
much smaller than the angular span of the caustic in the source frame.
 As we have
shown earlier, the rays close to the caustic are strongly bent by the lensing
effect, resulting in a significant shrinkage of the solid angle in the
Boyer-Lindquist coordinate. The small solid angle in the SMBH's frame also implies that the probability of seeing a highly
beamed BBH is low.  Far away from the direction of caustic, the magnification (or demagnification) factor remains
moderate. This result suggests that outside the caustic the lensing effect is
relatively mild.

The right panel of Figure~\ref{hdis} shows a much more complex pattern.  While
we still see a small region of strong magnification which is concentrated in
one direction, we also find large demagnification ($h_o/h\simeq0.1$) in a wide
range of directions.  This result suggests that the GWs originating inside the
caustic could be strongly lensed and demagnified.

Comparing the sky maps in both panels, we find that in many sky directions a
ray outside the caustic could coincide with a ray coming from inside the
caustic. This coincidence implies that an observer in such a special direction
can detect two images. Since these two images are emitted in very different
directions in the rest frame of the source, they will show different
magnification when detected by the observer.  Therefore, the distance inferred
from the two images will also differ, for the reason given in
Section~\ref{sec:massdis}.  This difference may further prevent us from finding
echoes that are physically associated with the same source
\citep[e.g.][]{kocsis13,gondan21,yu21}, in addition to the reason we have
pointed out in Section~\ref{sec:red}.

\begin{figure*}
    \centering
    \begin{tabular}{ccc}
        \includegraphics[width=0.33\textwidth]{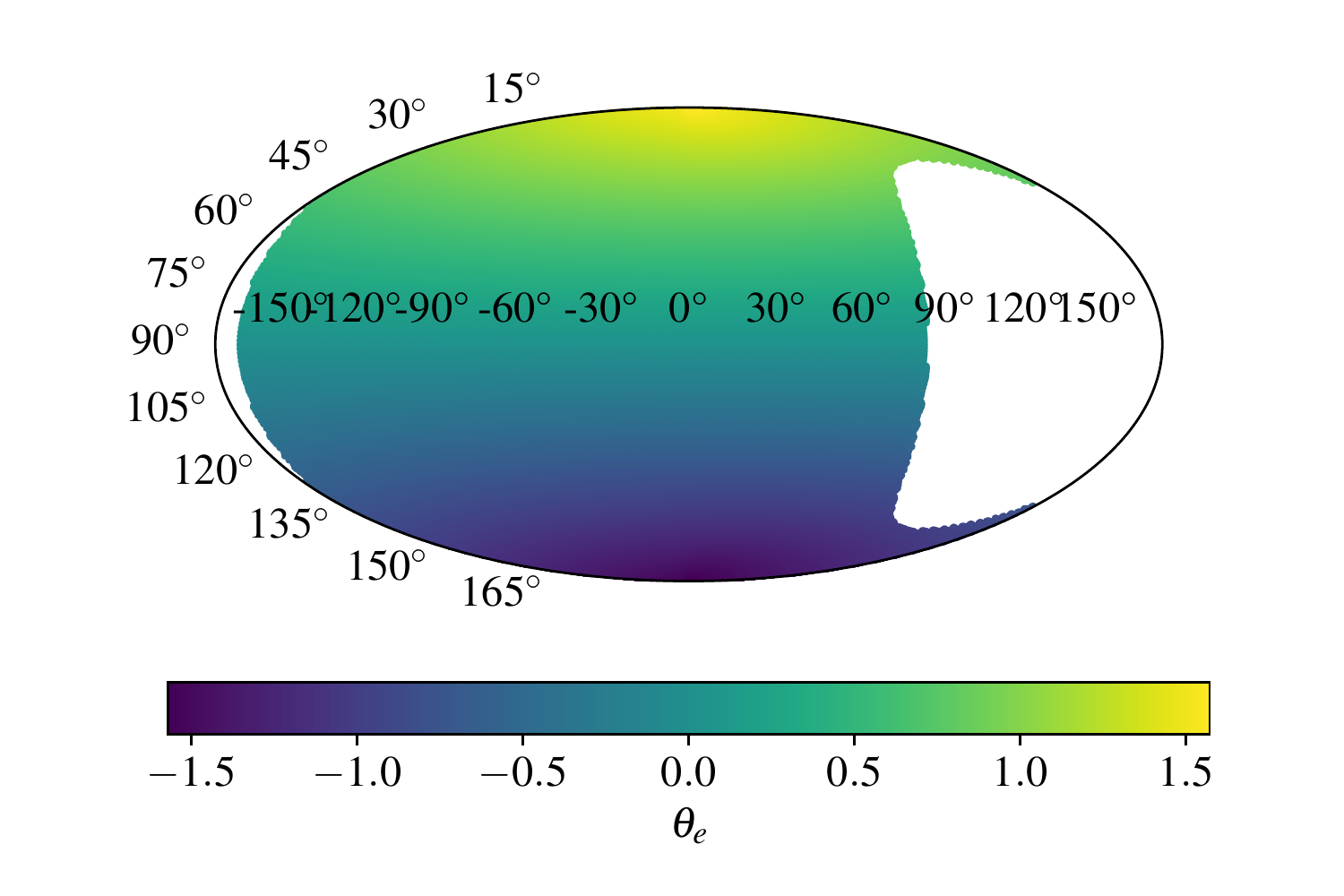}&
        \includegraphics[width=0.33\textwidth]{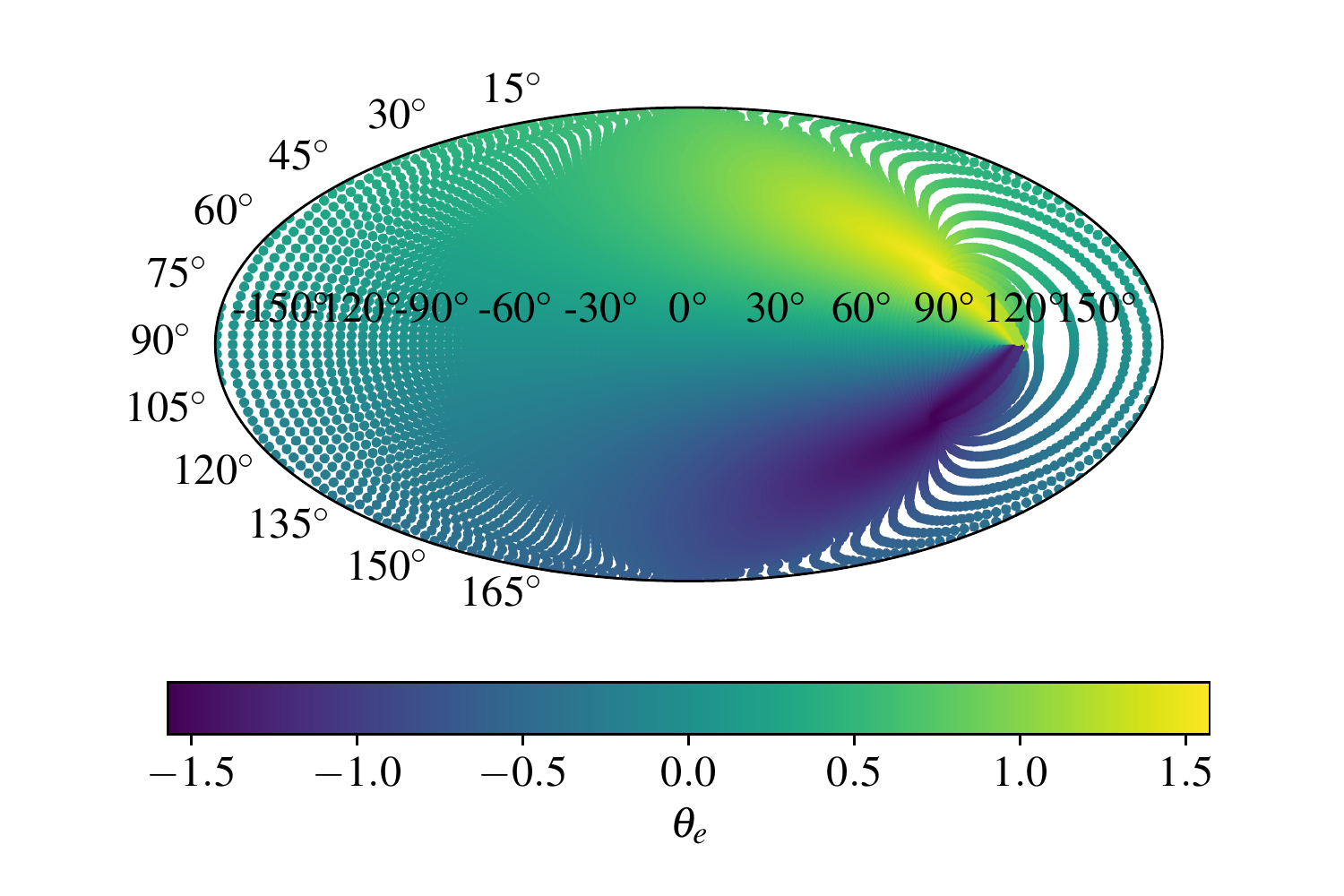}&
        \includegraphics[width=0.33\textwidth]{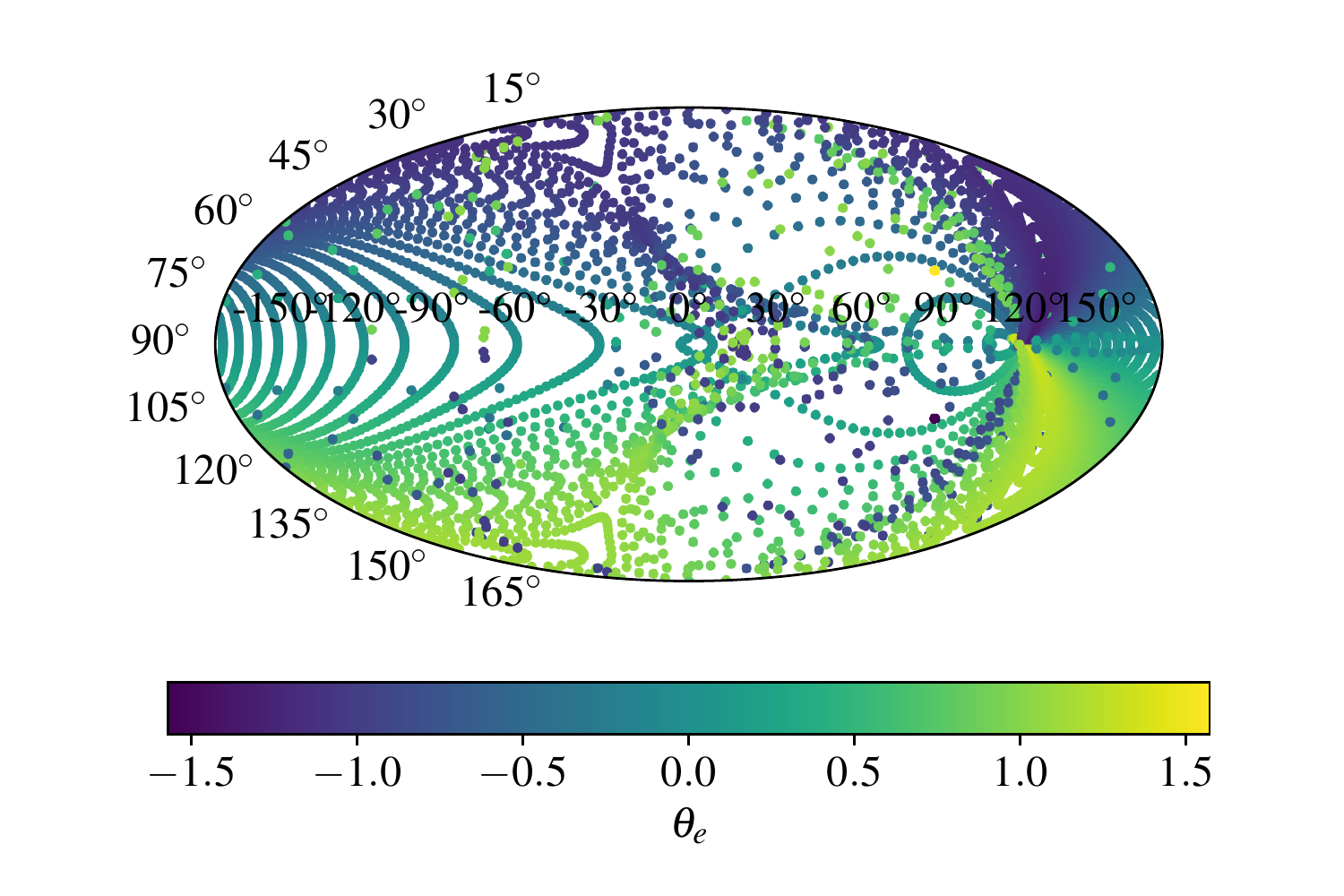}\\
        (a)&(b)&(c)
    \end{tabular} 
	\caption{Mapping the rays just emitted in the source frame (panel a) to
	the rays arriving at a distant observer (panels b and c).  Notice that in our
	model the rays arriving at a observer is viewed in the Boyer-Lindquist
	coordinates of the SMBH.  The color shows the inclination angle $\theta_e$ of the ray
	in the source frame. Panel (b) shows only those rays outside the
caustic and panel (c) shows the rays inside the caustic.} \label{geo}
\end{figure*}

To better understand the cause of the complexity of the pattern we have just
seen in the right panel of Figure~\ref{hdis}, we show in Figure~\ref{geo} a map
of the rays from the source frame (panel a) to the SMBH's frame (panels b
and c).  Here we use the inclination angle $\theta_e$ in the source frame to
color code the map. We also separate the rays originating outside the caustic
from those originating inside, and plot them in, respectively, panels (b) and
(c). 

First, by comparing panels (a) and (b), we find that the rays in the polar
region of the source frame are compressed towards the equatorial plane in the
SMBH's frame.This is partly a consequence of the beaming effect, and partly due to the lensing effect.
The comparison also shows that the rays originating outside the caustic do not cross
the equatorial plane, i.e., those from the northern (southern) hemisphere in
the source frame end up in the same hemisphere in the SMBH's frame.

Panel (c), which shows the rays inside the caustic, depicts a very different
picture.  Now the colors in the two hemispheres are largely reversed relative
to those in panel (a).  This result suggests that most of the rays inside the
caustic will cross the equatorial plane as they propagate to infinity. 

\begin{figure*}
    \centering
    \begin{tabular}{ccc}
        \includegraphics[width=0.33\textwidth]{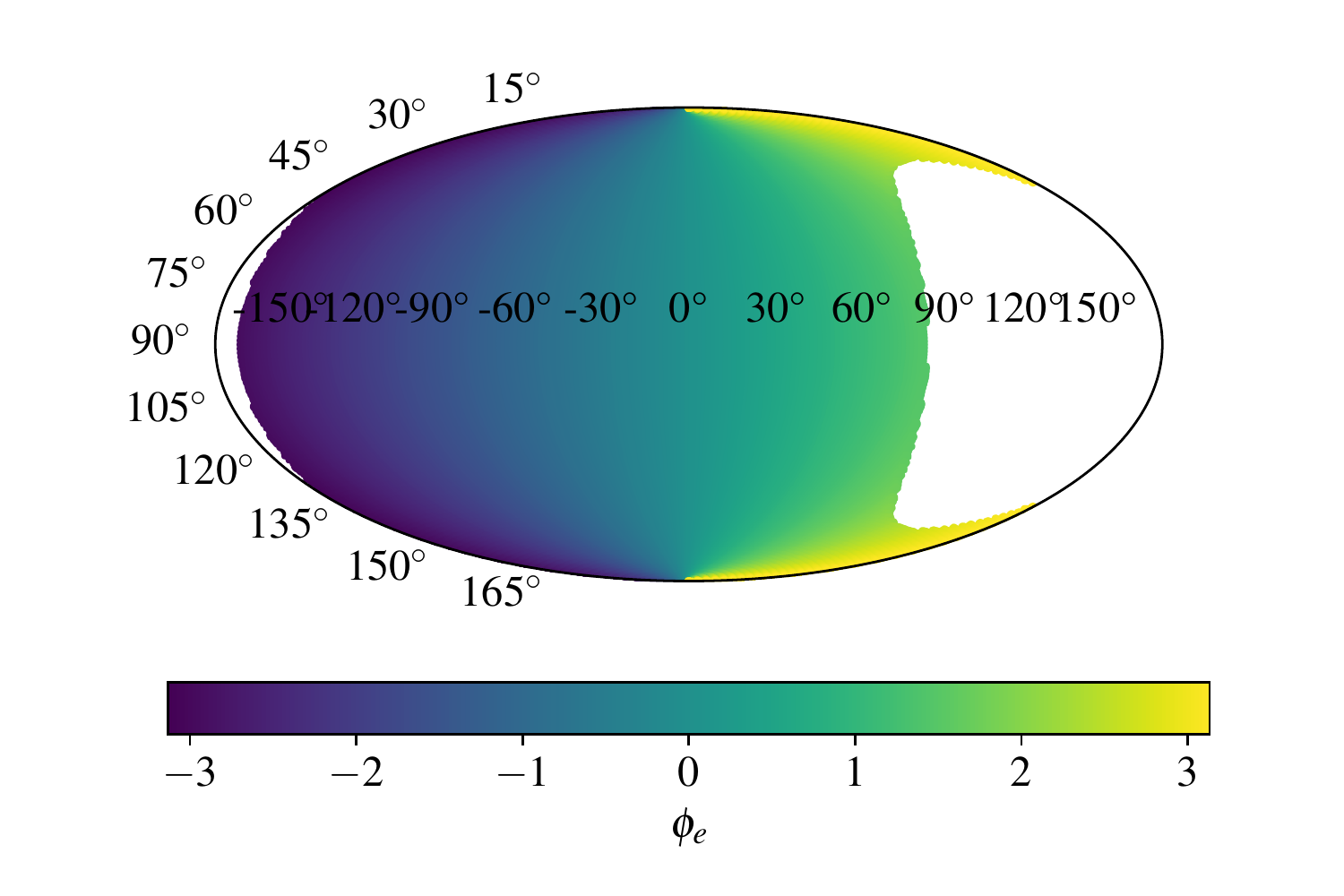}&
        \includegraphics[width=0.33\textwidth]{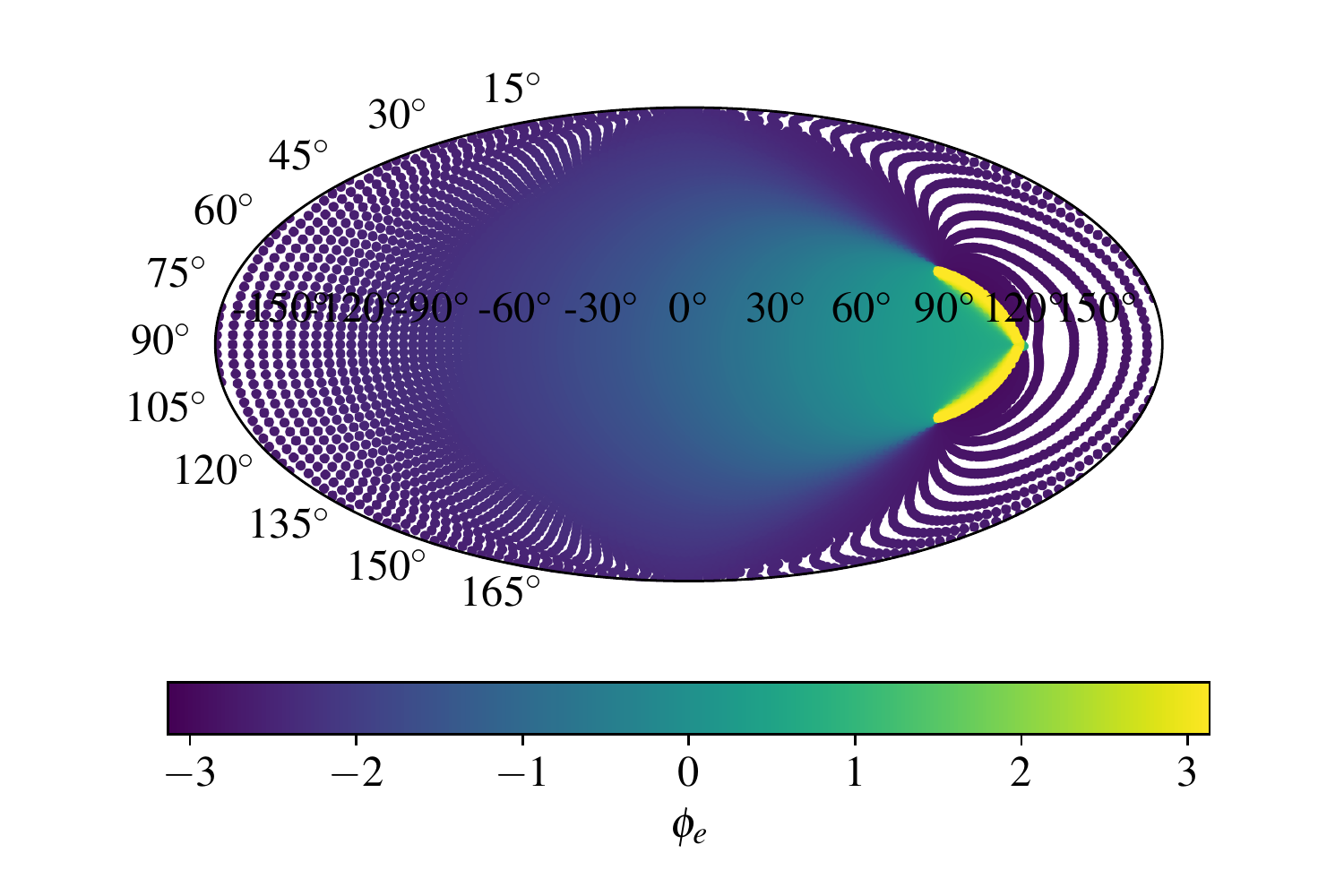}&
        \includegraphics[width=0.33\textwidth]{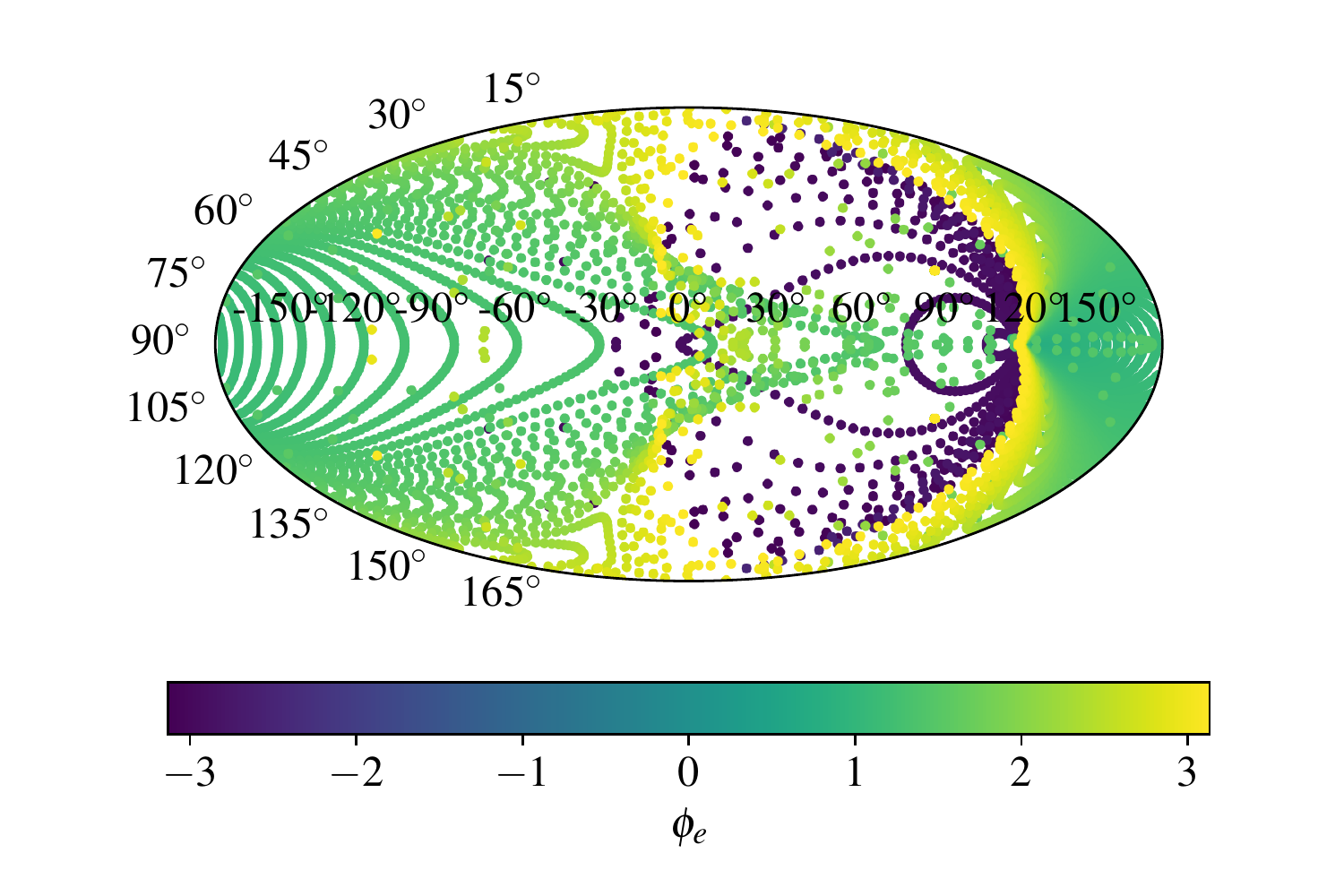}\\
        (a)&(b)&(c)
    \end{tabular}
	\caption{The same as Figure~\ref{geo} but color coded using the azimuthal angle
	$\phi_e$ in the source frame.}
    \label{geo1}
\end{figure*}

Figure~\ref{geo1} shows the same map but color coded by the azimuthal angle
$\phi_e$ in the rest frame of the source. We see a behavior similar to that in
Figure~\ref{geo}.  When viewed in the SMBH's frame, the rays outside the
caustic (panel b) exhibit a smooth transition in color, except for a high
concentration of bright yellow color in the direction of caustic.
The rays inside the caustic (panel c), on the contrary, show a much more
complicated pattern, which reflects the fact that they are crossing each other
in the azimuthal direction.

\subsection{Redshift as a function of viewing angle}\label{sec:zofangle}

\begin{figure*}
    \centering
    \begin{tabular}{ccc}
        \includegraphics[width=0.33\textwidth]{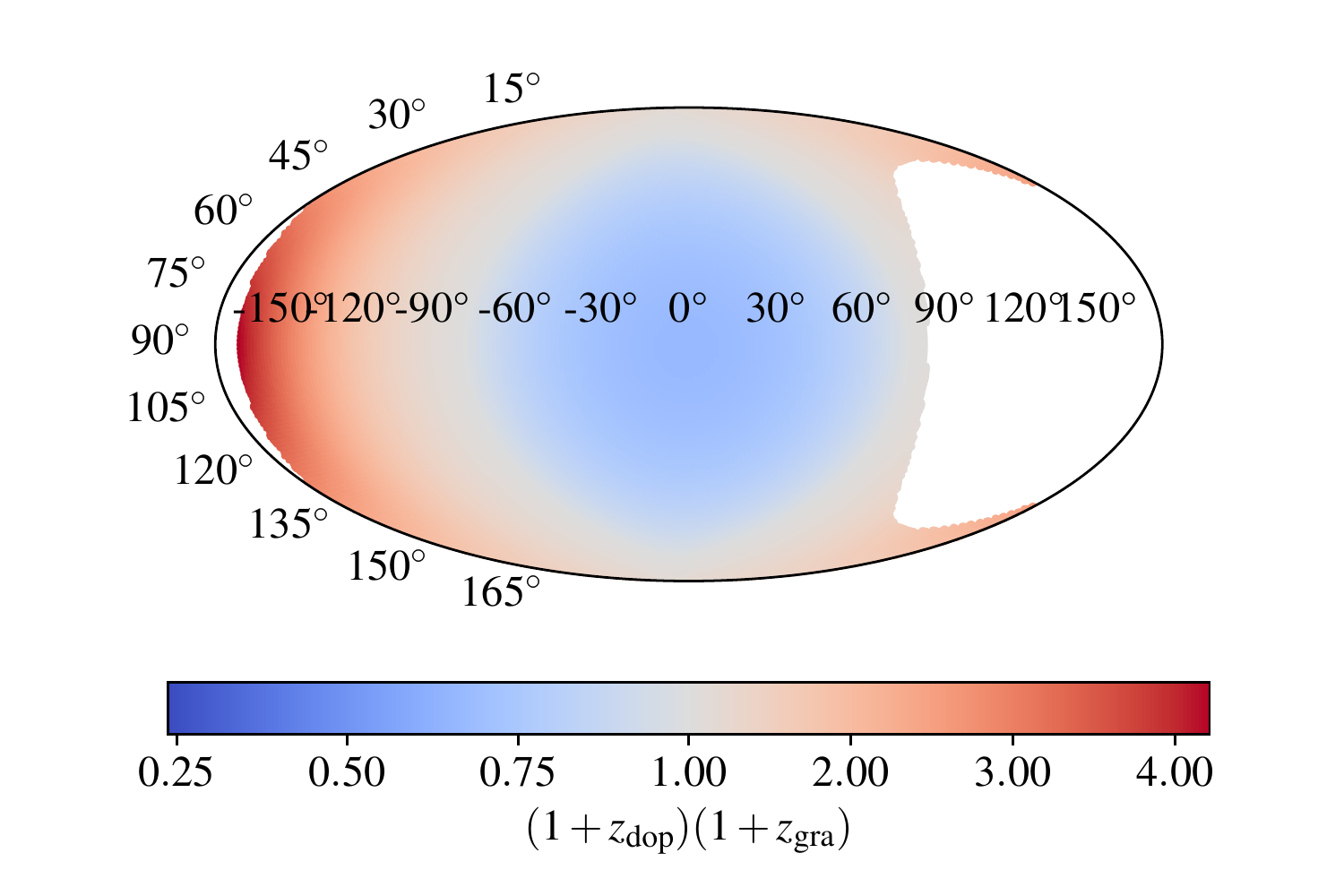}&
        \includegraphics[width=0.33\textwidth]{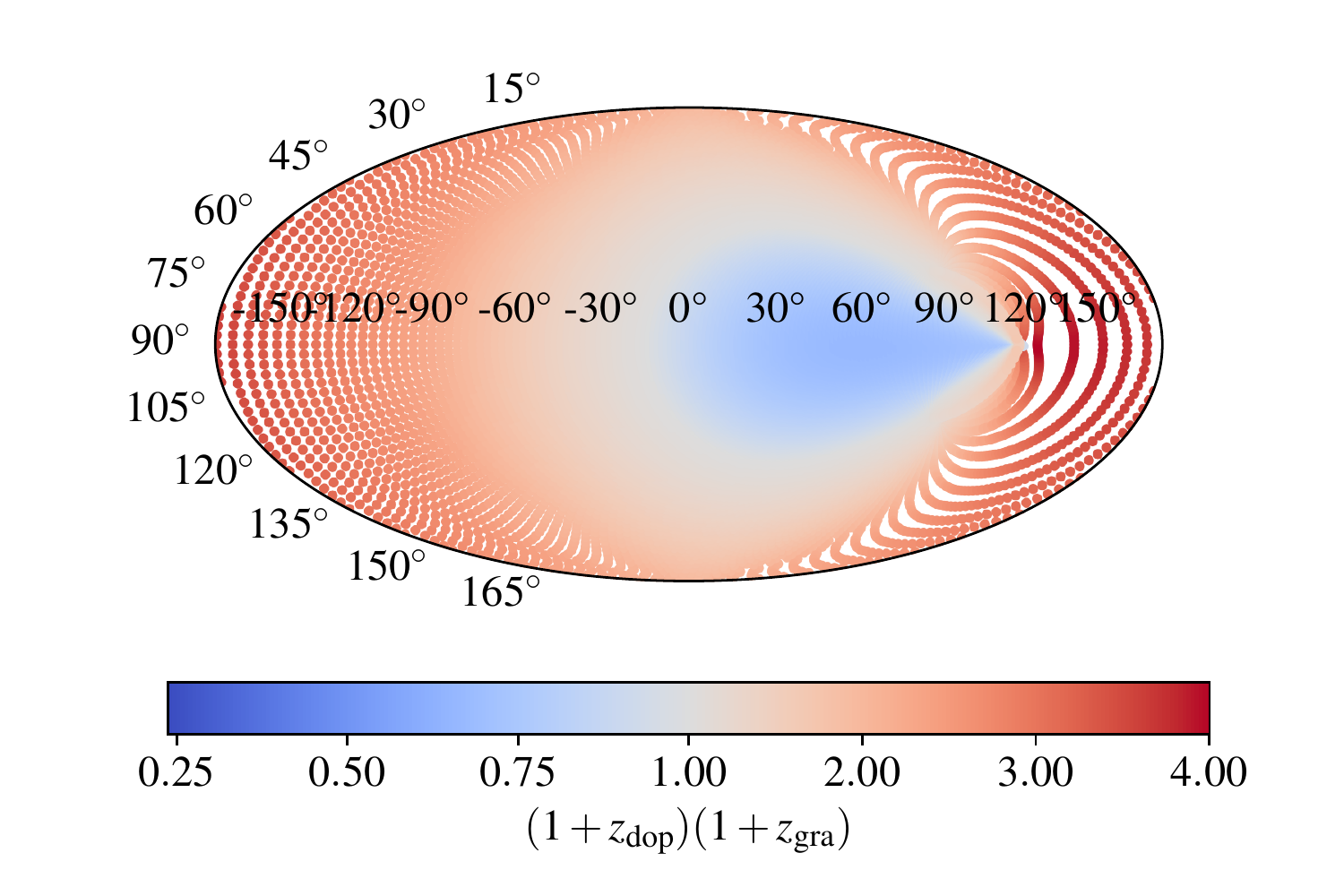}&
        \includegraphics[width=0.33\textwidth]{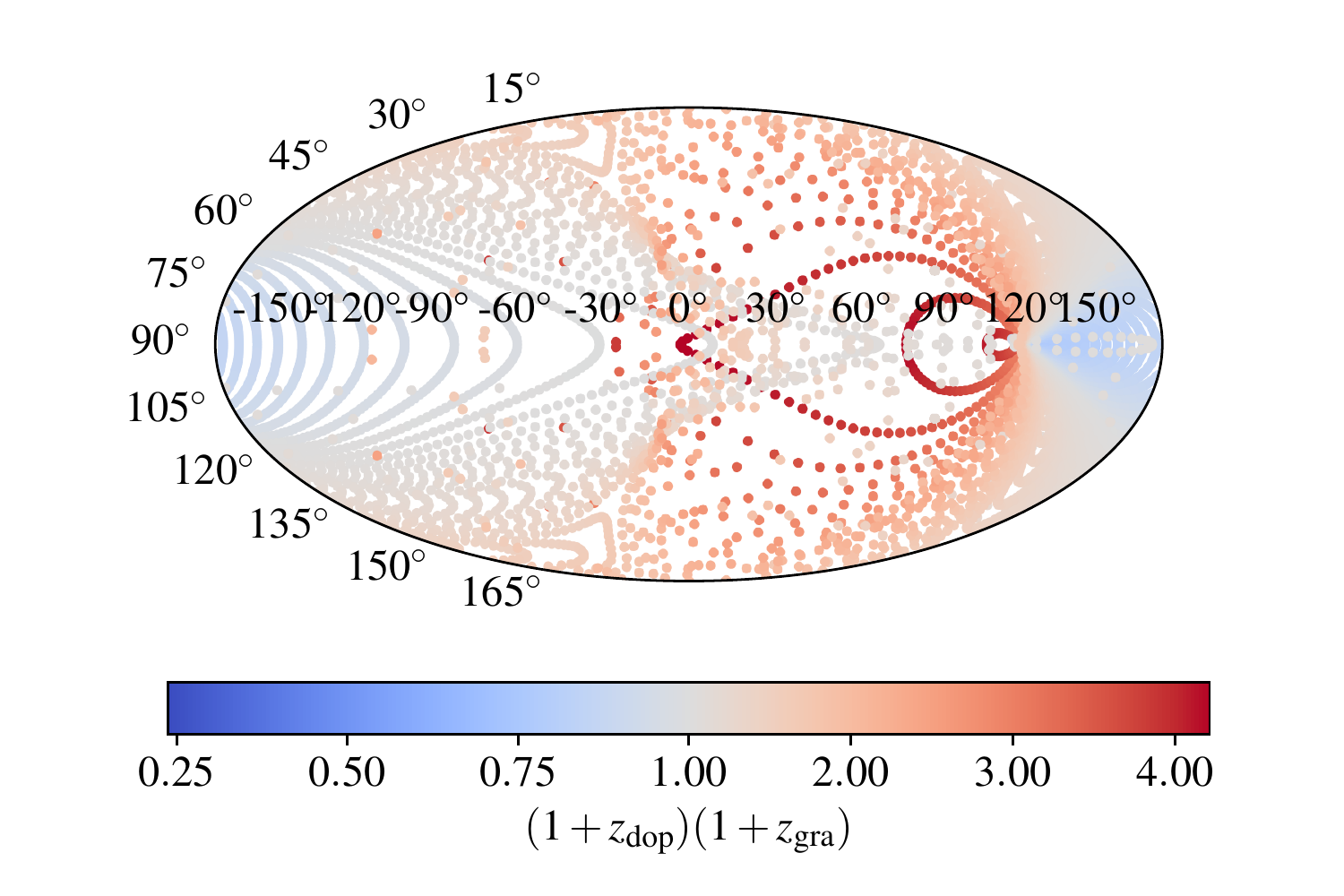}\\
        (a)&(b)&(c)
    \end{tabular}
	\caption{The same as Figure~\ref{geo} but color coded by the combined redshift 
	$1+z=(1+z_{\rm dop})(1+z_{\rm gra})$ due to
	the Dopper effect and the gravitational redshift.}
    \label{zdis}
\end{figure*}

The Doppler and gravitational redshifts of a ray seen by a distant observer
depend only on the rescaled constant $b$ (Section~\ref{sec:red}).
The combined redshift $(1+z_{\rm dop})(1+z_{\rm gra})$ is shown in
Figure~\ref{zdis}.  The parameters and the meaning of each panel are the same
as in Figure~\ref{geo}.

In panel (a), the rest frame of the source, we can see a Doppler blueshift of
the rays in the direction of the orbital velocity of the BBH.  Correspondingly,
the strongest redshift appears at $\phi_e\simeq180^\circ$, opposite to the
direction of the orbital velocity. Now we can see more clearly the asymmetry
between redshift and blueshift caused gravitational redshift, which we have
first mentioned in Section~\ref{sec:red}.  More specifically, in this example
the maximum redshift is $1+z\simeq 4.21$, while the maximum blueshift is about
$1+z\simeq0.67$.

In panel (b), which shows the observer's view of the rays outside the caustic,
we find that more than half of the sky is covered by redshift. This result
reinforces the asymmetry between redshift and blueshift, and corroborates our
earlier speculation that a BBH close to a SMBH is more likely to be seen
redshifted. Comparing this panel with the left one in Figure~\ref{hdis}, we
also find a close correlation between redshift and demagnification. Since
demagnification makes a BBH appear more distant (Section~\ref{sec:massdis}), we
conclude that BBHs close to SMBHs will more often appear heavier and further
away than they really are. It will be interesting to test such a positive
correlation between mass and distance in real GW observations.

Panel (c) shows the observer's view of the rays originating inside the
caustic.  Most of the rays are moderately redshifted or blueshifted, except for
a few direction in which the rays get highly redshifted. Again, we find that
the highly redshifted directions correspond to the directions of large
demagnification.  Observationally, this correspondence will produce a positive
correlation between the mass and the distance of a BBH. Moreover, by comparing
panels (b) and (c), we confirm the discovery made in Section~\ref{sec:red} that
in many direction the observer can detect two images with significantly
different redshifts. Comparing Figures~\ref{hdis} and \ref{zdis}, we can
see that outside the caustic, most rays which are moderately
shifted in frequency are also moderately magnified. However, it is no longer the case for
the rays inside the caustic, where many moderately Doppler shifted rays are
highly demagnified. The difference indicates that the two groups of rays 
will occupy different regions in the diagram of 
$1+z$ and $h_o/h$. Therefore, their apparent mass and apparent distance will also
follow different trends.
We will see this more clearly in the
following subsection.

\begin{figure}
    \includegraphics[width=0.5\textwidth]{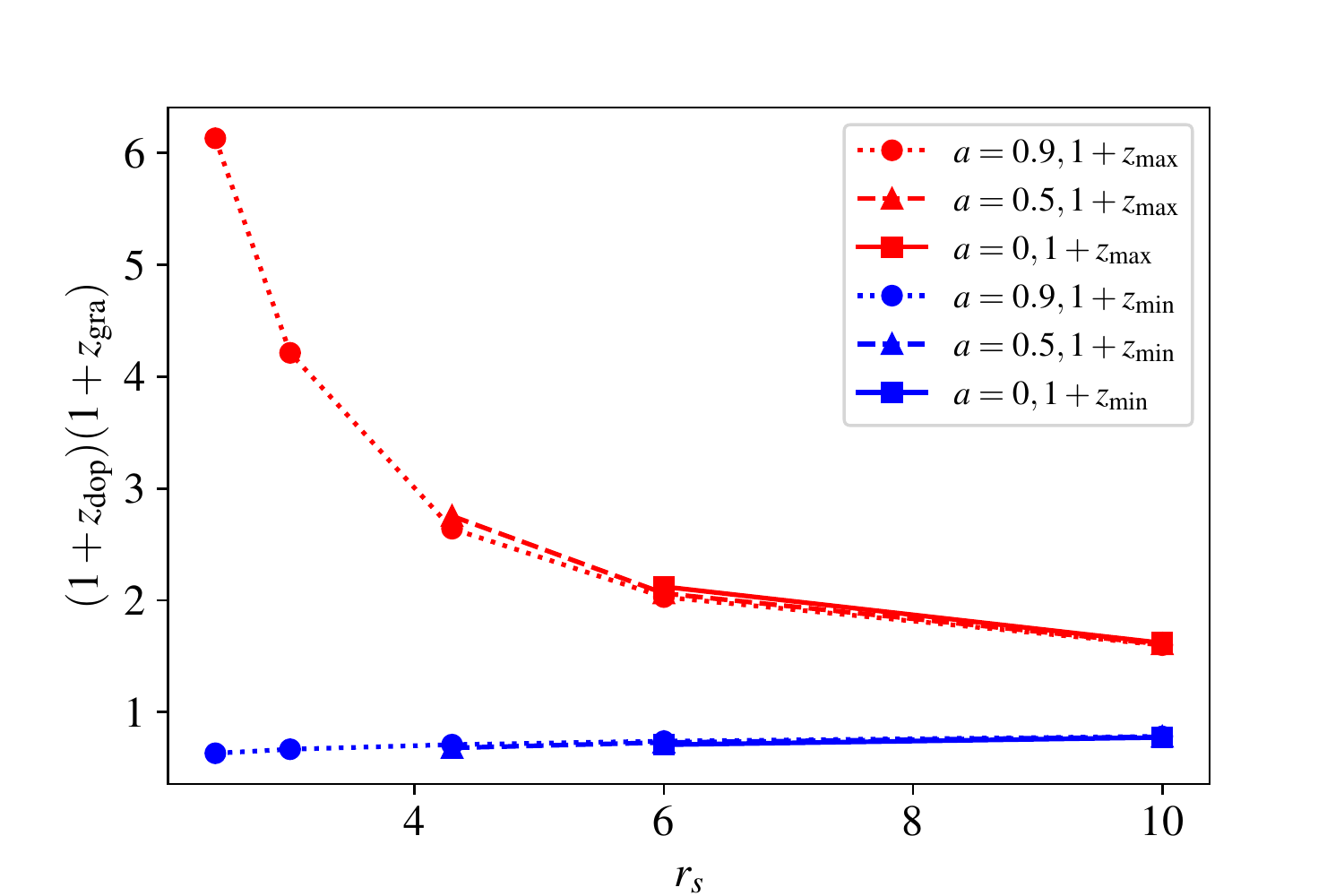}
	\caption{Dependence of the maximum redshift (red symbols) and maximum blueshift (blue symbols) 
	on the radius of the source $r_s$ and the spin $a$ of the SMBH. }
    \label{zdis2}
\end{figure}

So far we have only considered the system with the parameters $r_s=3M$ and
$a=0.9$.  To see the dependence of the combined redshift $1+z$ on these
parameters, we show in Figure~\ref{zdis2} the maximum and minimum values of
$1+z$ as a function of $r_s$ and $a$. Notice that $1+z<1$ corresponds to a
blueshift of GW, and the radius of the innermost stable circular orbit (ISCO)
depends on $a$.  We find that the value of $1+z$ is insensitive to the spin
paramiter $a$, but more sensitive to the distance $r_s$ between the BBH and the
SMBH. The dependence on $r_s$ is more prominent for redshifted rays (red
symbols).  When $r_s=(2.4, 3, 4.3, 6, 10)M$, the maximum redshift is
$1+z_{\rm{max}}\simeq(6.10,4.21,2.64,2.02,1.60)$. These values suggest
that a stellar-mass BH similar to those found in X-ray binaries (e.g., with a
typical mass of $10M_\odot$) will appear significantly more massive (e.g.,
$20-60M_\odot$) in the rest frame of a GW detector. Interestingly, the apparent
mass in this example seems to be consistent with the massive BHs detected by
LIGO and Virgo.

\section{Appearance  of a BBH close to a SMBH}\label{sec:app}

In the previous sections, we have shown how a nearby SMBH affects the
redshift and amplitude of the GW emitted by a BBH. In this subsection, we will
further study the impact on the chirp mass $\mathcal{M}_o$ and apparent
distance $d_o$ in the detector's frame. We will also investigate the effects on
the inferred parameters of the BBH, namely the inferred chirp mass
$\mathcal{M}'$ and the inferred redshift $z_{d_o}$. The latter two parameters
are related to $\mathcal{M}_o$ and $h_o$ through Equations (\ref{eq:dodL}) and
(\ref{eq:Mrelation}).  

\begin{figure}
    \includegraphics[width=0.5\textwidth]{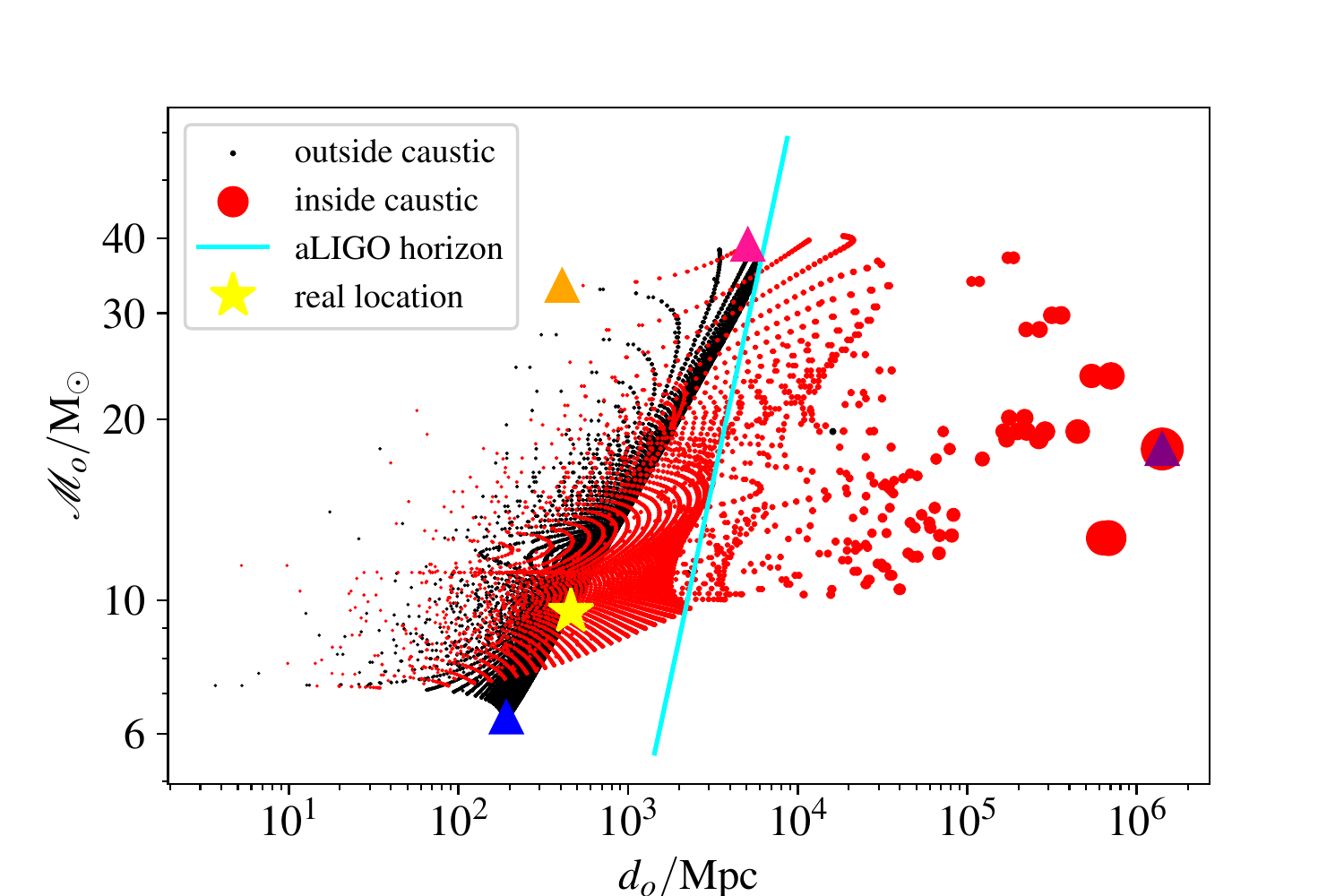}
\caption{Distance and chirp mass of a BBH in the detector frame (see
Section~\ref{sec:massdis} for explanation).
The dots correspond to those rays shown in Figure~\ref{hdis}. Here we have also assumed
that $m_1=m_2=10M_\odot$, and the SMBH resides at
a cosmological redshift of $z_{\rm cos}=0.1$.
The black dots correspond to the rays originating outside the caustic and the red ones inside the caustic.
The size of a dot is proportional to the solid angle of the ray bundle in the Boyer-Lindquist frame, 
and hence is directly correlated with the probability of detecting the ray bundle.
The yellow star marks the location of the BBH if there is no SMBH around the BBH.
The cyan line shows the detection horizon of advanced LIGO (aLIGO\protect\footnotemark).
The four triangles are four representative images, which will be analyzed later to
illustrate the dependence of the apparent mass and disntance on the real cosmological
redshift of the source.
}
    \label{zm1}
\end{figure}

In Figure~\ref{zm1} we show the distribution of the rays emitted by one BBH in
the plane of $d_o$ and $\mathcal{M}_o$. The black dots represent the rays
outside the caustic and the red ones inside the caustic. The size of a dot
represents the solid angle that the ray bundles span in the SMBH's sky. 
Therefore, it is proportional to the possibility for an observer in an
arbitrary direction to see the signal. 
Generally speaking, the
the dots at larger $\mathcal{M}_o$ correspond to the 
rays that are more redshifted.  Again, we can see the asymmetry between 
redshift and blueshift, since the majority of the dots lie above the 
real chirp mass (marked by the yellow star) of the BBH.
Moreover, the dots at larger $d_o$ could also come from the rays that are
significantly demagnified, since $d_o$ is
proportional to the solid angle
according to Equations~(\ref{eq:dodL}) and (\ref{eq:ho}).
This effect produces those big red dots which occupy the right-half of the plot.
We notice that the combined redshift $1+z$ is normally smaller than $10$ in our examples,
but the apparent distance $d_o$ for some rays can be more than ten times greater than the 
real luminosity distance $d_L$ of the source.
This discrepancy indicates that the large apparent distances are caused not by the redshift effect,
but the lensing of the GWs by the SMBH.

\footnotetext{\url{https://range.ligo.org/}}

If we focus on the rays outside the caustic (black dots), we find that they
are aligned in the diagonal direction. Such a positive correlation between
apparent mass and distance is caused by the fact that both parameters are
positively correlated with the redshift (see Section~\ref{sec:massdis}).  The
rays originating outside the caustic (red dots) show a more complex
distribution. In particular, there are many branches and they occupy a much
larger parameter space. The overall shape of the distribution is also quite
different from that of the black dots. This difference reflects the transition
of the relationship between redshift and magnification as the rays enter the
caustic, which we have already discussed in Section~\ref{sec:zofangle}.
Nevertheless, the positively correlation between apparent mass and apparent
distance is still present.

\begin{figure}
\includegraphics[width=0.5\textwidth]{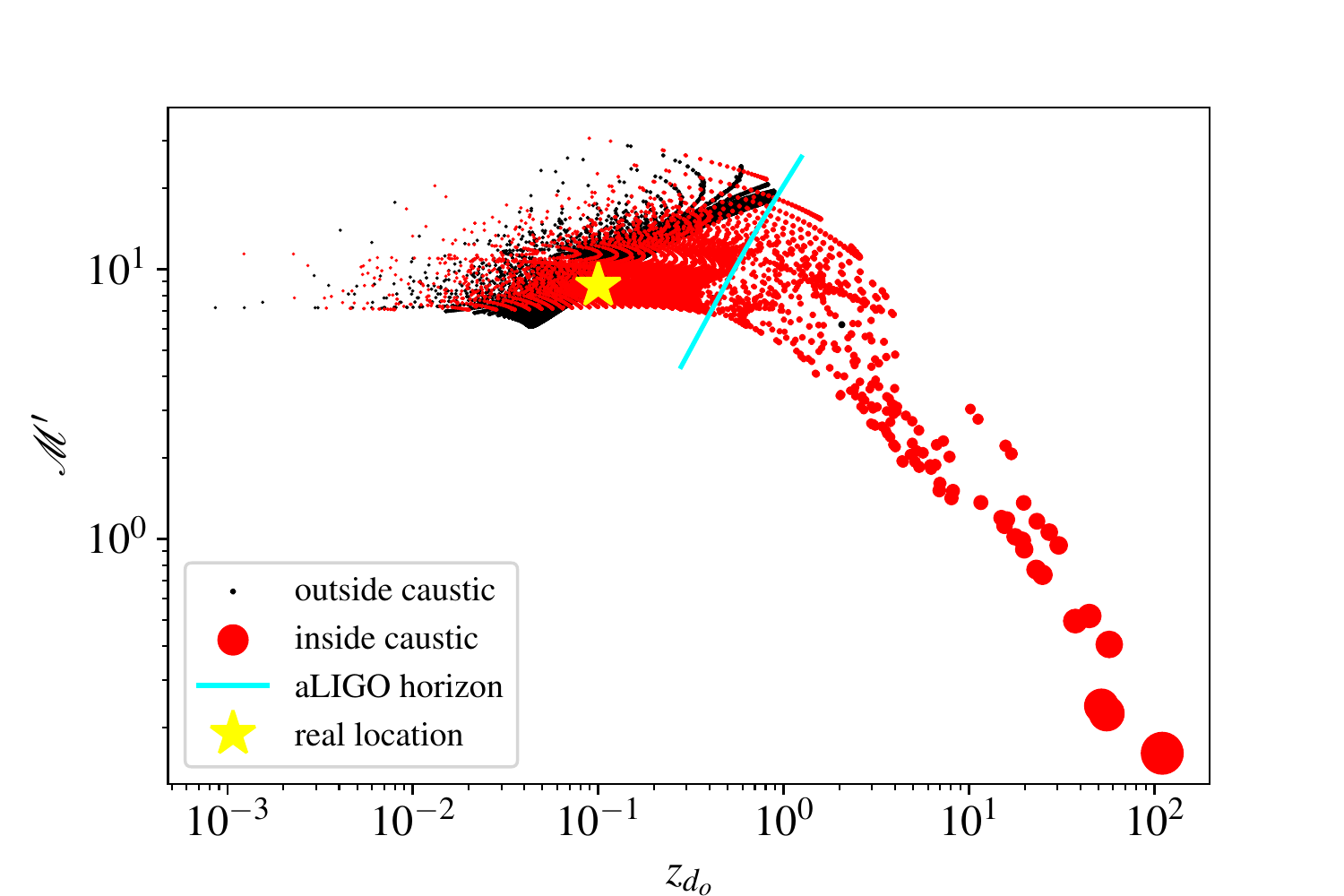}
\caption{Apparent cosmological redshift and apparent chirp mass of a BBH inferred by an observer who
does not know that the BBH resides close to a SMBH. The dots and their sizes have the same meanings
as in Figure~\ref{zm1}. The yellow star marks the real location of the BBH.
The cyan solid line indicates the detection horizon of the advanced LIGO detectors.
}
    \label{zm2}
\end{figure}

Figure~\ref{zm2} shows the distribution of the rays in the
$z_{d_o}$-$\mathcal{M}'$ plane. Comparing with Figure~\ref{zm1}, we first
notice that the maximum $\mathcal{M}'$ is significantly smaller than the
maximum $\mathcal{M}_o$. The decrement is caused by the factor of $1+z_{d_o}$,
as we have shown in Equation~(\ref{eq:Mprime}).  We also notice that the
distribution of the dots are much flatter in Figure~\ref{zm2} than in
Figure~\ref{zm1}. In particular, the big red dots, which correspond to the rays
that are significantly demagnified, bend over and trace the off diagonal in
Figure~\ref{zm2}.  The cause is the same as before, due to the correlation
between demagnification and larger $1+z_{d_o}$.  Interestingly, the biggest red
dots occupy a region of large apparent redshift and small (sub-solar) BH mass.
Such a region was thought to be occupied only by primordial black holes \citep{2018PhRvL.121w1103A}.  Unfortunately, the BBHs in this region are below the current
detection limit. 

\begin{figure}
    \includegraphics[width=0.5\textwidth]{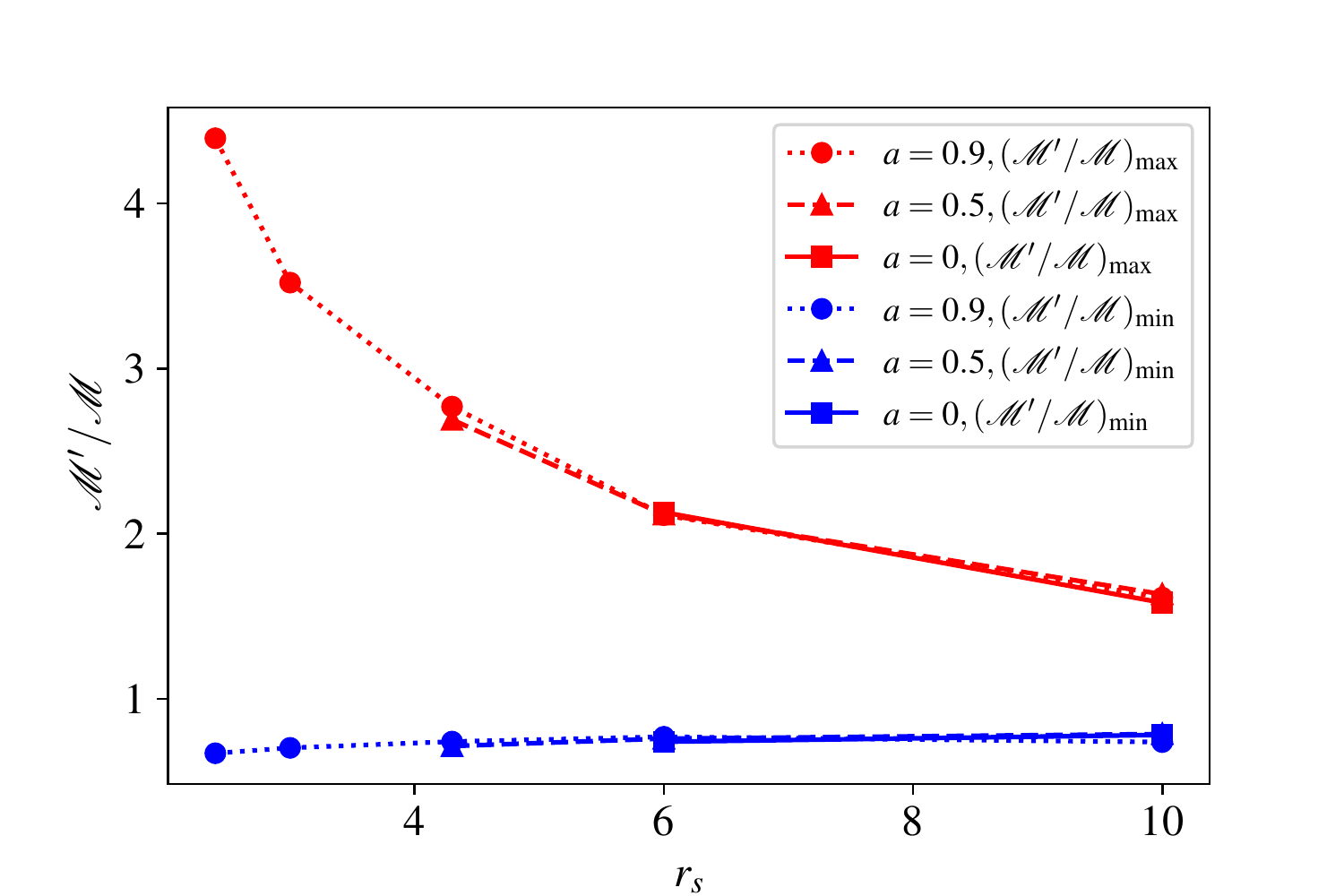}
    \caption{Dependence of the inferred mass on $r_s$ and $a$. 
Red dots show the maximum values of $\mathcal{M}'/\mathcal{M}$ at the corresponding $r_s$, 
and the blue dots show the minimum values of $\mathcal{M}'/\mathcal{M}$. 
We only include those rays detectable by aLIGO. The values of $r_s$ are $(2.4, 3, 4.3, 6, 10)M$. 
	The real cosmological redshift of the source is set to $z_{\mathrm{cos}}=0.1$.}
    \label{mdis}
\end{figure}

In Figure~\ref{mdis} we show the range of $\mathcal{M}'/\mathcal{M}$ as a function
of $r_s$ and $a$. Notice that when determining the minimum value of
$\mathcal{M}'/\mathcal{M}$, we have excluded the dots below the detection limit
of aLIGO. Similar to Figure~\ref{zdis2}, the result is more sensitive to $r_s$
than to $a$. 
In this work we take $z_{\rm{cos}}=0.1$, and hence when $r_s=(2.4, 3, 4.3, 6,
10)M$ and $a=0.9$, the maximum values of $\mathcal{M}'/\mathcal{M}$ are
$(4.39,3.52,2.77,2.11,1.61)$.  The corresponding minimum values are $(0.67,0.70,0.74,0.77,0.81)$.
These values indicate that an observer unaware of the presence of the SMBH
would overestimate (underestimate) the mass of the BBHs by a factor of $4.39$
($0.67$).

\begin{figure}
    \includegraphics[width=0.5\textwidth]{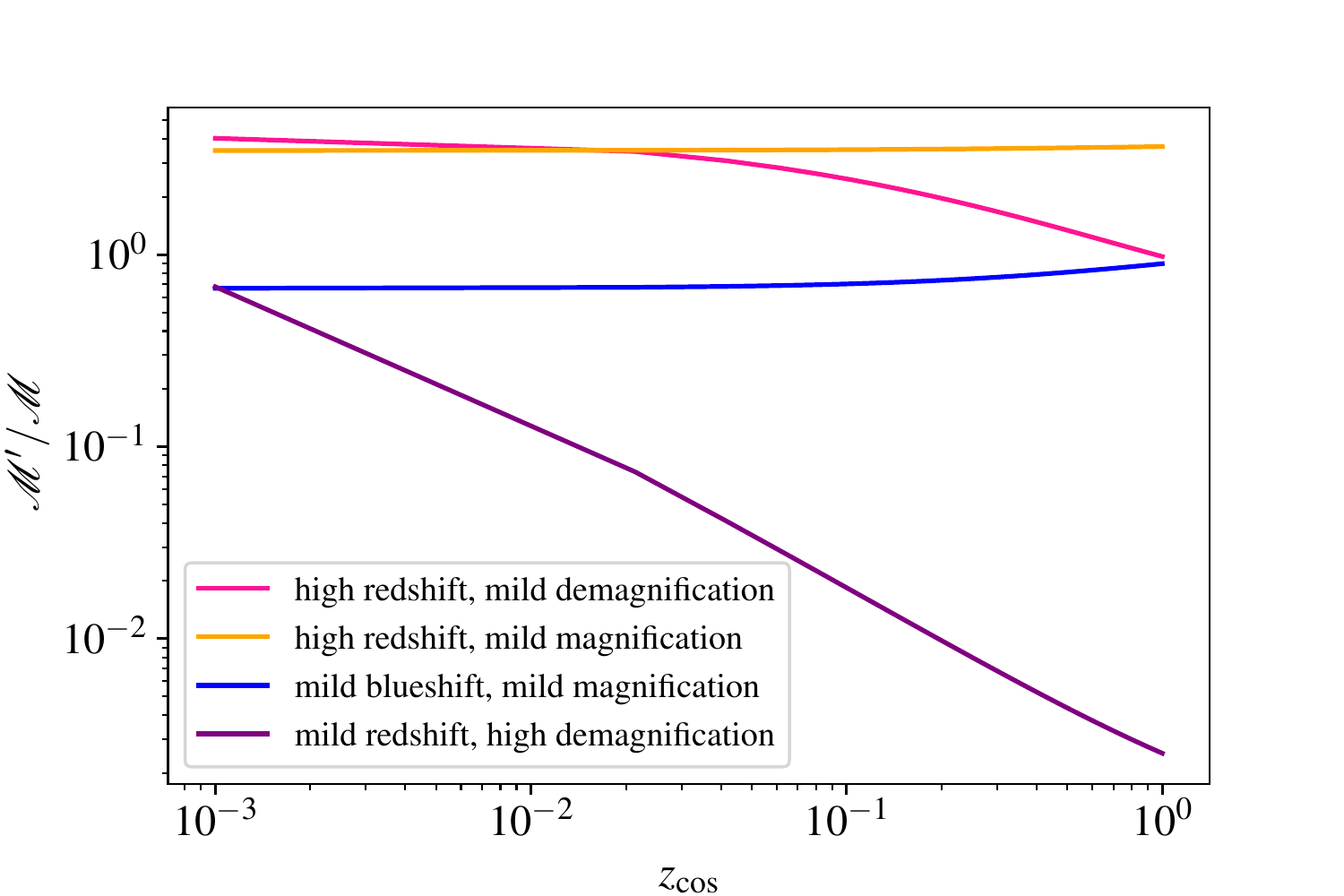}
    \caption{Dependence of $\mathcal{M}'/\mathcal{M}$ on the 
real cosmological redshift of the source $z_{\rm{cos}}$. 
The four lines correspond to the four images marked by triangles in Figure~\ref{zm1}.
The paramters other than $z_{\rm{cos}}$ are the same as in Figure~\ref{zm1}.
    }
    \label{zz}
\end{figure}

We notice that the value of $\mathcal{M}'/\mathcal{M}$ depends on the choice of
the real cosmological redshift $z_{\rm{cos}}$ of the source.  The dependence is
nonlinear due to the nonlinear relationship between $1+z_{\rm{cos}}$ and
$1+z_{d_o}$.  To illustrate the relationship, we show four examples in
Figure~\ref{zz}. They are chosen from Figure~\ref{zm1}, where the value of
$z_{\rm{cos}}$ was fixed at $0.1$. Now we allow $z_{\rm{cos}}$ to change and
study the variation of $\mathcal{M}'/\mathcal{M}$.
We find that if a ray gets magnified (orange and blue lines), the $\mathcal{M}'/\mathcal{M}$
of the image will slightly increase with $z_{\rm{cos}}$. 
In this case, an ignorant observer may further overestimate the mass of the BBH if the
system resides at higher cosmological redshift.
For demagnified images (red and purple), the relationship reverses.
Therefore, a BBH at lower cosmological redshift could appear more massive. In this case,
we find that the maximum value of $\mathcal{M}'/\mathcal{M}$ 
in our previous example becomes $5.61$ if we use $z_{\rm{cos}}=0.01$.

\section{Discussion}\label{sec:dis}

In this work we have studied how the apparent mass and distance of a GW source
(a BBH) would be affected by a nearby Kerr SMBH.  Our study is motivated by the
recent theoretical discovery that BBHs could merge within tens of gravitational
radii of a SMBH \citep{chen_han_2018,addison_gracia-linares_2019,peng21}.  We
showed that the appearance of the BBH depends on the frequency shift and
(de)magnification of the GW signal.  By analyzing the null geodesics
originating within $10$ gravitational radii of the SMBH, we found a higher
probability for the GW signal to appear redshifted and demagnified, rather than
blueshifted and magnified, when detected by a distant observer.  Such an
asymmetry indicates that the observer, unaware of the presence of the SMBH, is
more likely to overestimate the mass and distance of the BBH.

Our examples suggest that a BBH residing at approximately the ISCO
($r_s\simeq2.4$) of a Kerr SMBH ($a=0.9$) could appear $4.4-5.6$ times more
massive than its real mass, if the systems resides at a cosmological redshift
between  $z_{\rm{cos}}=0.01$ and $0.1$. For this reason, a binary composed of
normal stellar-mass BHs of $(10-20)M_\odot$ would appear to contain overly
massive BHs, with a mass of $(40-110)M_\odot$.  Interestingly, such massive BHs
have been detected by LIGO/Virgo. For example, the rest mass of the primary BH
of GW190521 is estimated to be $m_1=95.3^{+28.7}_{18.9}$, and that of
GW190426\_190642 is about $m_1=106.9^{+41.6}_{25.2}$
\citep{2021arXiv210801045T}. It is still too early to conclude that 
GW190521 and GW190426\_190642 are coming from the vicinities of SMBHs, since
conventional models can also explain their existence \citep{2020ApJ...903L...5R,2021PhRvL.126t1101B,2021PhRvL.126e1101D,2020ApJ...902L..26F}. 
However, the fact, that out of the $\sim90$ BBHs detected so far by LIGO/Virgo two
contain BHs of $\sim100M_\odot$, is roughly consistent with the estimation
that about $1\%$ of the LIGO/Virgo BBHs are produced at the inner edges of the
accretion disks in AGNs \citep{peng21}.

One major difference between our work and the earlier ones on the magnification
of GWs by SMBHs \citep[e.g.][]{kocsis13,gondan21,yu21} is that we studied the
angular dependence of the gravitational lensing effect. Therefore, we found
that in the majority of the directions the GWs will be, in fact, demagnified.
We have shown that demagnified images could be misinterpreted as subsolar-mass
BHs merging at $z_{\rm{cos}}>10$ (Figure~\ref{zm2}), if the data analysis does
not account for the presence of a nearby SMBH. Such events could be
misidentified as primordial BHs produced by density fluctuation in the
early universe \citep[e.g.][]{2018PhRvL.121w1103A,2020JCAP...08..039C}.
Although these events fall below the sensitivity of the current detectors, they
may be found by future ground-based detectors such as the Einstein Telescope
and Cosmic Explorer \citep{2010CQGra..27s4002P,2017CQGra..34d4001A}.

It is known that an impulsive light source close to a SMBH can produce at least
two images relative to a distant observer, a primary image corresponding to the
ray going directly from the source to the observer and a secondary one caused
by the ray going around the SMBH  \citep[e.g.][]{thompson19}. The same applies
to a impulsive GW source, as we have shown in Figures~\ref{fig:localz},
\ref{hdis} and \ref{zdis}. We found that the two images will be redshifted and
(de)magnified differently, because in the source frame they are emitted in two
different directions and around the SMBH they follow different null geodesics.
The difference in redshift and (de)magnified, according to the analysis in
Section~\ref{sec:massdis}, will result in different apparent mass and distance
for the two images. Therefore, an observer may interpret them as two physically
separated, independent events.

Additional informational may help the observer realize that two events seemly
uncorrelated by mass and distance may actually be the two images of the same
BBH merger. First, the sky locations of the two images should be the same
since the two images originate from the same galaxy.  Second, there is a
typical time delay of $\sim10M\simeq1.4M_8$ hours between the two images, where
$M_8\equiv M/(10^8M_\odot)$.  The duration  corresponds to the light crossing
time over the region of strong lensing around the SMBH. Third, the two images
should yield the same mass ratio $m_2/m_1$ for the BBH, since frequency shift
and the lensing effect do not affect the measurement of this parameter (as long
as the impulsive approximation of the signal is valid). Therefore, looking for
correlated events in the three-dimensional space of sky location, signal
arrival time and mass ratio could eventually reveal those BBHs close to SMBHs.

The impulsive approximation adopted by our model would break down if the SMBH
is less massive than $10^5M_\odot$. In this case, the size of the ISCO is of
the order of $10^5-10^6$ km and the BBH would have traversed a distance of
$\sim c\times0.1s\simeq3\times10^4$ km in the typical duration ($0.1$ s) of the
signal.  The latter is a substantial fraction of the former, indicating that
the the velocity and the position of the BBH relative to the distant observer
would have varied substantially during the time span of the signal.  This
variation effectively changes the view angle of the source by the observer.  A
change of the viewing angle will change the frequency and amplitude of the GW
signal, as we have discussed extensively in Section~\ref{sec:results}.  In
addition, the plus and cross polarizations of the signal will also change
because they also depend on the viewing angle.  

We do not yet know whether the
above effects could provide a potential signature for the observer to
distinguish the BBHs around SMBHs from other isolated ones, because the effects
related to a varying viewing angle can be mimicked by a precession of the orbit
induced by, e.g., the spin of the BHs or the orbital eccentricity of the binary
\citep{2004PhRvD..70d2001V,2021PhRvL.126t1101B}. Breaking such a degeneracy requires
more careful modeling of the waveform of a BBH moving within a few gravitational
radii of a Kerr SMBH \citep[e.g.][]{cardoso21}.

\section*{Acknowledgements}

This work is supported by the National Key Research and Development Program of
China Grant No. 2021YFC2203002 and the National Natural Science Foundation of
China (NSFC) grant No. 11991053.  The computation in this work was performed on
the High Performance Computing Platform of the Centre for Life Science, Peking
University.

\section*{Data Availability}

The data underlying this article will be shared on reasonable request to the corresponding author.






\bibliographystyle{mnras.bst}
\bibliography{draft2,aamnem99,biblio,bibbase,redshift}

\begin{thebibliography}{}
\makeatletter
\relax
\def\mn@urlcharsother{\let\do\@makeother \do\$\do\&\do\#\do\^\do\_\do\%\do\~}
\def\mn@doi{\begingroup\mn@urlcharsother \@ifnextchar [ {\mn@doi@}
  {\mn@doi@[]}}
\def\mn@doi@[#1]#2{\def\@tempa{#1}\ifx\@tempa\@empty \href
  {http://dx.doi.org/#2} {doi:#2}\else \href {http://dx.doi.org/#2} {#1}\fi
  \endgroup}
\def\mn@eprint#1#2{\mn@eprint@#1:#2::\@nil}
\def\mn@eprint@arXiv#1{\href {http://arxiv.org/abs/#1} {{\tt arXiv:#1}}}
\def\mn@eprint@dblp#1{\href {http://dblp.uni-trier.de/rec/bibtex/#1.xml}
  {dblp:#1}}
\def\mn@eprint@#1:#2:#3:#4\@nil{\def\@tempa {#1}\def\@tempb {#2}\def\@tempc
  {#3}\ifx \@tempc \@empty \let \@tempc \@tempb \let \@tempb \@tempa \fi \ifx
  \@tempb \@empty \def\@tempb {arXiv}\fi \@ifundefined
  {mn@eprint@\@tempb}{\@tempb:\@tempc}{\expandafter \expandafter \csname
  mn@eprint@\@tempb\endcsname \expandafter{\@tempc}}}

\bibitem[\protect\citeauthoryear{{Abbott} et~al.,}{{Abbott}
  et~al.}{2017}]{2017CQGra..34d4001A}
{Abbott} B.~P.,  et~al., 2017, \mn@doi [Classical and Quantum Gravity]
  {10.1088/1361-6382/aa51f4}, \href
  {https://ui.adsabs.harvard.edu/abs/2017CQGra..34d4001A} {34, 044001}

\bibitem[\protect\citeauthoryear{{Abbott} et~al.,}{{Abbott}
  et~al.}{2018}]{2018PhRvL.121w1103A}
{Abbott} B.~P.,  et~al., 2018, \mn@doi [\prl] {10.1103/PhysRevLett.121.231103},
  \href {https://ui.adsabs.harvard.edu/abs/2018PhRvL.121w1103A} {121, 231103}

\bibitem[\protect\citeauthoryear{{Addison}, {Gracia-Linares}, {Laguna}  \&
  {Larson}}{{Addison} et~al.}{2019}]{addison_gracia-linares_2019}
{Addison} E.,  {Gracia-Linares} M.,  {Laguna} P.,   {Larson} S.~L.,  2019,
  \mn@doi [General Relativity and Gravitation] {10.1007/s10714-019-2523-4},
  \href {https://ui.adsabs.harvard.edu/abs/2019GReGr..51...38A} {51, 38}

\bibitem[\protect\citeauthoryear{{Antonini} \& {Perets}}{{Antonini} \&
  {Perets}}{2012}]{antonini_perets_2012}
{Antonini} F.,  {Perets} H.~B.,  2012, \mn@doi [\apj]
  {10.1088/0004-637X/757/1/27}, \href
  {http://adsabs.harvard.edu/abs/2012ApJ...757...27A} {757, 27}

\bibitem[\protect\citeauthoryear{{Arca Sedda}}{{Arca
  Sedda}}{2020}]{ArcaSedda20}
{Arca Sedda} M.,  2020, \mn@doi [\apj] {10.3847/1538-4357/ab723b}, \href
  {https://ui.adsabs.harvard.edu/abs/2020ApJ...891...47A} {891, 47}

\bibitem[\protect\citeauthoryear{{Bardeen}}{{Bardeen}}{1970}]{1970ApJ...162...71B}
{Bardeen} J.~M.,  1970, \mn@doi [\apj] {10.1086/150635}, \href
  {https://ui.adsabs.harvard.edu/abs/1970ApJ...162...71B} {162, 71}

\bibitem[\protect\citeauthoryear{{Bardeen}, {Press}  \& {Teukolsky}}{{Bardeen}
  et~al.}{1972}]{1972ApJ...178..347B}
{Bardeen} J.~M.,  {Press} W.~H.,   {Teukolsky} S.~A.,  1972, \mn@doi [\apj]
  {10.1086/151796}, \href
  {https://ui.adsabs.harvard.edu/abs/1972ApJ...178..347B} {178, 347}

\bibitem[\protect\citeauthoryear{{Bartos}, {Kocsis}, {Haiman}  \&
  {M{\'a}rka}}{{Bartos} et~al.}{2017}]{bartos_kocsis_2017}
{Bartos} I.,  {Kocsis} B.,  {Haiman} Z.,   {M{\'a}rka} S.,  2017, \mn@doi
  [\apj] {10.3847/1538-4357/835/2/165}, \href
  {http://adsabs.harvard.edu/abs/2017ApJ...835..165B} {835, 165}

\bibitem[\protect\citeauthoryear{{Baruteau}, {Cuadra}  \& {Lin}}{{Baruteau}
  et~al.}{2011}]{baruteau11}
{Baruteau} C.,  {Cuadra} J.,   {Lin} D.~N.~C.,  2011, \mn@doi [\apj]
  {10.1088/0004-637X/726/1/28}, \href
  {https://ui.adsabs.harvard.edu/abs/2011ApJ...726...28B} {726, 28}

\bibitem[\protect\citeauthoryear{{Bonvin}, {Caprini}, {Sturani}  \&
  {Tamanini}}{{Bonvin} et~al.}{2017}]{bonvin_caprini_2017}
{Bonvin} C.,  {Caprini} C.,  {Sturani} R.,   {Tamanini} N.,  2017, \mn@doi
  [\prd] {10.1103/PhysRevD.95.044029}, \href
  {http://adsabs.harvard.edu/abs/2017PhRvD..95d4029B} {95, 044029}

\bibitem[\protect\citeauthoryear{{Bozza}}{{Bozza}}{2008}]{bozza08}
{Bozza} V.,  2008, \mn@doi [\prd] {10.1103/PhysRevD.78.103005}, \href
  {https://ui.adsabs.harvard.edu/abs/2008PhRvD..78j3005B} {78, 103005}

\bibitem[\protect\citeauthoryear{{Bozza} \& {Mancini}}{{Bozza} \&
  {Mancini}}{2004}]{bozza04}
{Bozza} V.,  {Mancini} L.,  2004, \mn@doi [\apj] {10.1086/422309}, \href
  {https://ui.adsabs.harvard.edu/abs/2004ApJ...611.1045B} {611, 1045}

\bibitem[\protect\citeauthoryear{{Bustillo}, {Sanchis-Gual}, {Torres-Forn{\'e}}
   \& {Font}}{{Bustillo} et~al.}{2021}]{2021PhRvL.126t1101B}
{Bustillo} J.~C.,  {Sanchis-Gual} N.,  {Torres-Forn{\'e}} A.,   {Font} J.~A.,
  2021, \mn@doi [\prl] {10.1103/PhysRevLett.126.201101}, \href
  {https://ui.adsabs.harvard.edu/abs/2021PhRvL.126t1101B} {126, 201101}

\bibitem[\protect\citeauthoryear{{Campbell} \& {Matzner}}{{Campbell} \&
  {Matzner}}{1973}]{campbell73}
{Campbell} G.~A.,  {Matzner} R.~A.,  1973, \mn@doi [Journal of Mathematical
  Physics] {10.1063/1.1666159}, \href
  {https://ui.adsabs.harvard.edu/abs/1973JMP....14....1C} {14, 1}

\bibitem[\protect\citeauthoryear{{Cardoso}, {Duque}  \& {Khanna}}{{Cardoso}
  et~al.}{2021}]{cardoso21}
{Cardoso} V.,  {Duque} F.,   {Khanna} G.,  2021, \mn@doi [\prd]
  {10.1103/PhysRevD.103.L081501}, \href
  {https://ui.adsabs.harvard.edu/abs/2021PhRvD.103h1501C} {103, L081501}

\bibitem[\protect\citeauthoryear{{Carter}}{{Carter}}{1968}]{1968PhRv..174.1559C}
{Carter} B.,  1968, \mn@doi [Physical Review] {10.1103/PhysRev.174.1559}, \href
  {https://ui.adsabs.harvard.edu/abs/1968PhRv..174.1559C} {174, 1559}

\bibitem[\protect\citeauthoryear{{Chen}}{{Chen}}{2021}]{chen21book}
{Chen} X.,  2021, in , Handbook of Gravitational Wave Astronomy.
p.~39, \mn@doi{10.1007/978-981-15-4702-7_39-1}

\bibitem[\protect\citeauthoryear{{Chen} \& {Han}}{{Chen} \&
  {Han}}{2018}]{chen_han_2018}
{Chen} X.,  {Han} W.-B.,  2018, \mn@doi [Communications Physics]
  {10.1038/s42005-018-0053-0}, \href
  {https://ui.adsabs.harvard.edu/abs/2018CmPhy...1...53C} {1, 53}

\bibitem[\protect\citeauthoryear{{Chen} \& {Huang}}{{Chen} \&
  {Huang}}{2020}]{2020JCAP...08..039C}
{Chen} Z.-C.,  {Huang} Q.-G.,  2020, \mn@doi [\jcap]
  {10.1088/1475-7516/2020/08/039}, \href
  {https://ui.adsabs.harvard.edu/abs/2020JCAP...08..039C} {2020, 039}

\bibitem[\protect\citeauthoryear{{Chen} \& {Zhang}}{{Chen} \&
  {Zhang}}{2022}]{chen22gem}
{Chen} X.,  {Zhang} Z.,  2022, \mn@doi [\prd] {10.1103/PhysRevD.106.103040},
  \href {https://ui.adsabs.harvard.edu/abs/2022PhRvD.106j3040C} {106, 103040}

\bibitem[\protect\citeauthoryear{{Chen}, {Li}  \& {Cao}}{{Chen}
  et~al.}{2019}]{chen_li_2019}
{Chen} X.,  {Li} S.,   {Cao} Z.,  2019, \mn@doi [Monthly Notices of the Royal
  Astronomical Society: Letters] {10.1093/mnrasl/slz046}, \href
  {http://adsabs.harvard.edu/abs/2017arXiv170310543C} {485, L141}

\bibitem[\protect\citeauthoryear{{Corral-Santana}, {Casares},
  {Mu{\~n}oz-Darias}, {Bauer}, {Mart{\'\i}nez-Pais}  \&
  {Russell}}{{Corral-Santana} et~al.}{2016}]{2016A&A...587A..61C}
{Corral-Santana} J.~M.,  {Casares} J.,  {Mu{\~n}oz-Darias} T.,  {Bauer} F.~E.,
  {Mart{\'\i}nez-Pais} I.~G.,   {Russell} D.~M.,  2016, \mn@doi [\aap]
  {10.1051/0004-6361/201527130}, \href
  {https://ui.adsabs.harvard.edu/abs/2016A&A...587A..61C} {587, A61}

\bibitem[\protect\citeauthoryear{{Cunningham} \& {Bardeen}}{{Cunningham} \&
  {Bardeen}}{1973}]{cunningham73}
{Cunningham} C.~T.,  {Bardeen} J.~M.,  1973, \mn@doi [\apj] {10.1086/152223},
  \href {https://ui.adsabs.harvard.edu/abs/1973ApJ...183..237C} {183, 237}

\bibitem[\protect\citeauthoryear{{D'Orazio} \& {Loeb}}{{D'Orazio} \&
  {Loeb}}{2020}]{dorazio20}
{D'Orazio} D.~J.,  {Loeb} A.,  2020, \mn@doi [\prd]
  {10.1103/PhysRevD.101.083031}, \href
  {https://ui.adsabs.harvard.edu/abs/2020PhRvD.101h3031D} {101, 083031}

\bibitem[\protect\citeauthoryear{{De Luca}, {Desjacques}, {Franciolini}, {Pani}
   \& {Riotto}}{{De Luca} et~al.}{2021}]{2021PhRvL.126e1101D}
{De Luca} V.,  {Desjacques} V.,  {Franciolini} G.,  {Pani} P.,   {Riotto} A.,
  2021, \mn@doi [\prl] {10.1103/PhysRevLett.126.051101}, \href
  {https://ui.adsabs.harvard.edu/abs/2021PhRvL.126e1101D} {126, 051101}

\bibitem[\protect\citeauthoryear{{Ford} \& {McKernan}}{{Ford} \&
  {McKernan}}{2021}]{ford21mckernan}
{Ford} K.~E.~S.,  {McKernan} B.,  2021, arXiv e-prints, \href
  {https://ui.adsabs.harvard.edu/abs/2021arXiv210903212F} {p. arXiv:2109.03212}

\bibitem[\protect\citeauthoryear{{Fragione}, {Grishin}, {Leigh}, {Perets}  \&
  {Perna}}{{Fragione} et~al.}{2019}]{fragione19}
{Fragione} G.,  {Grishin} E.,  {Leigh} N. W.~C.,  {Perets} H.~B.,   {Perna} R.,
   2019, \mn@doi [\mnras] {10.1093/mnras/stz1651}, \href
  {https://ui.adsabs.harvard.edu/abs/2019MNRAS.488...47F} {488, 47}

\bibitem[\protect\citeauthoryear{{Fragione}, {Loeb}  \& {Rasio}}{{Fragione}
  et~al.}{2020}]{2020ApJ...902L..26F}
{Fragione} G.,  {Loeb} A.,   {Rasio} F.~A.,  2020, \mn@doi [\apjl]
  {10.3847/2041-8213/abbc0a}, \href
  {https://ui.adsabs.harvard.edu/abs/2020ApJ...902L..26F} {902, L26}

\bibitem[\protect\citeauthoryear{{Gond{\'a}n} \& {Kocsis}}{{Gond{\'a}n} \&
  {Kocsis}}{2022}]{gondan21}
{Gond{\'a}n} L.,  {Kocsis} B.,  2022, \mn@doi [\mnras]
  {10.1093/mnras/stac1985}, \href
  {https://ui.adsabs.harvard.edu/abs/2022MNRAS.515.3299G} {515, 3299}

\bibitem[\protect\citeauthoryear{{Gralla} \& {Lupsasca}}{{Gralla} \&
  {Lupsasca}}{2020}]{2020PhRvD.101d4032G}
{Gralla} S.~E.,  {Lupsasca} A.,  2020, \mn@doi [\prd]
  {10.1103/PhysRevD.101.044032}, \href
  {https://ui.adsabs.harvard.edu/abs/2020PhRvD.101d4032G} {101, 044032}

\bibitem[\protect\citeauthoryear{{Gralla}, {Lupsasca}  \&
  {Strominger}}{{Gralla} et~al.}{2018}]{gralla18}
{Gralla} S.~E.,  {Lupsasca} A.,   {Strominger} A.,  2018, \mn@doi [\mnras]
  {10.1093/mnras/sty039}, \href
  {https://ui.adsabs.harvard.edu/abs/2018MNRAS.475.3829G} {475, 3829}

\bibitem[\protect\citeauthoryear{{Gr{\"o}bner}, {Ishibashi}, {Tiwari}, {Haney}
  \& {Jetzer}}{{Gr{\"o}bner} et~al.}{2020}]{grobner20}
{Gr{\"o}bner} M.,  {Ishibashi} W.,  {Tiwari} S.,  {Haney} M.,   {Jetzer} P.,
  2020, \mn@doi [\aap] {10.1051/0004-6361/202037681}, \href
  {https://ui.adsabs.harvard.edu/abs/2020A&A...638A.119G} {638, A119}

\bibitem[\protect\citeauthoryear{{Hasse}, {Kriele}  \& {Perlick}}{{Hasse}
  et~al.}{1996}]{1996CQGra..13.1161H}
{Hasse} W.,  {Kriele} M.,   {Perlick} V.,  1996, \mn@doi [Classical and Quantum
  Gravity] {10.1088/0264-9381/13/5/027}, \href
  {https://ui.adsabs.harvard.edu/abs/1996CQGra..13.1161H} {13, 1161}

\bibitem[\protect\citeauthoryear{{Holz} \& {Hughes}}{{Holz} \&
  {Hughes}}{2005}]{holz_hughes_2005}
{Holz} D.~E.,  {Hughes} S.~A.,  2005, \mn@doi [\apj] {10.1086/431341}, \href
  {http://adsabs.harvard.edu/abs/2005ApJ...629...15H} {629, 15}

\bibitem[\protect\citeauthoryear{{Igata}, {Kohri}  \& {Ogasawara}}{{Igata}
  et~al.}{2021}]{2021PhRvD.103j4028I}
{Igata} T.,  {Kohri} K.,   {Ogasawara} K.,  2021, \mn@doi [\prd]
  {10.1103/PhysRevD.103.104028}, \href
  {https://ui.adsabs.harvard.edu/abs/2021PhRvD.103j4028I} {103, 104028}

\bibitem[\protect\citeauthoryear{{Inayoshi}, {Tamanini}, {Caprini}  \&
  {Haiman}}{{Inayoshi} et~al.}{2017}]{inayoshi_tamanini_2017}
{Inayoshi} K.,  {Tamanini} N.,  {Caprini} C.,   {Haiman} Z.,  2017, \mn@doi
  [\prd] {10.1103/PhysRevD.96.063014}, \href
  {https://ui.adsabs.harvard.edu/abs/2017PhRvD..96f3014I} {96, 063014}

\bibitem[\protect\citeauthoryear{Isaacson}{Isaacson}{1967}]{isaacson_1967}
Isaacson R.~A.,  1967, \mn@doi [Phys. Rev.] {10.1103/PhysRev.166.1263}, 166,
  1263

\bibitem[\protect\citeauthoryear{{Kocsis}}{{Kocsis}}{2013}]{kocsis13}
{Kocsis} B.,  2013, \mn@doi [\apj] {10.1088/0004-637X/763/2/122}, \href
  {https://ui.adsabs.harvard.edu/abs/2013ApJ...763..122K} {763, 122}

\bibitem[\protect\citeauthoryear{{LIGO Scientific Collaboration} \& {Virgo
  Collaboration}}{{LIGO Scientific Collaboration} \& {Virgo
  Collaboration}}{2016}]{2016ApJ...818L..22A}
{LIGO Scientific Collaboration} {Virgo Collaboration} 2016, \mn@doi [\apjl]
  {10.3847/2041-8205/818/2/L22}, \href
  {https://ui.adsabs.harvard.edu/abs/2016ApJ...818L..22A} {818, L22}

\bibitem[\protect\citeauthoryear{{LIGO Scientific Collaboration} \& {Virgo
  Collaboration}}{{LIGO Scientific Collaboration} \& {Virgo
  Collaboration}}{2020}]{2020ApJ...900L..13A}
{LIGO Scientific Collaboration} {Virgo Collaboration} 2020, \mn@doi [\apjl]
  {10.3847/2041-8213/aba493}, \href
  {https://ui.adsabs.harvard.edu/abs/2020ApJ...900L..13A} {900, L13}

\bibitem[\protect\citeauthoryear{{Lawrence}}{{Lawrence}}{1973}]{lawrence73}
{Lawrence} J.~K.,  1973, \mn@doi [\prd] {10.1103/PhysRevD.7.2275}, \href
  {https://ui.adsabs.harvard.edu/abs/1973PhRvD...7.2275L} {7, 2275}

\bibitem[\protect\citeauthoryear{{Leigh} et~al.,}{{Leigh}
  et~al.}{2018}]{2018MNRAS.474.5672L}
{Leigh} N.~W.~C.,  et~al., 2018, \mn@doi [\mnras] {10.1093/mnras/stx3134},
  \href {https://ui.adsabs.harvard.edu/abs/2018MNRAS.474.5672L} {474, 5672}

\bibitem[\protect\citeauthoryear{{Levin} \& {Perez-Giz}}{{Levin} \&
  {Perez-Giz}}{2008}]{2008PhRvD..77j3005L}
{Levin} J.,  {Perez-Giz} G.,  2008, \mn@doi [\prd]
  {10.1103/PhysRevD.77.103005}, \href
  {https://ui.adsabs.harvard.edu/abs/2008PhRvD..77j3005L} {77, 103005}

\bibitem[\protect\citeauthoryear{{McClintock}, {Narayan}  \&
  {Steiner}}{{McClintock} et~al.}{2014}]{mcclintock14}
{McClintock} J.~E.,  {Narayan} R.,   {Steiner} J.~F.,  2014, \mn@doi [\ssr]
  {10.1007/s11214-013-0003-9}, \href
  {http://adsabs.harvard.edu/abs/2014SSRv..183..295M} {183, 295}

\bibitem[\protect\citeauthoryear{{McKernan}, {Ford}, {Lyra}  \&
  {Perets}}{{McKernan} et~al.}{2012}]{mckernan_ford_2012}
{McKernan} B.,  {Ford} K.~E.~S.,  {Lyra} W.,   {Perets} H.~B.,  2012, \mn@doi
  [\mnras] {10.1111/j.1365-2966.2012.21486.x}, \href
  {https://ui.adsabs.harvard.edu/abs/2012MNRAS.425..460M} {425, 460}

\bibitem[\protect\citeauthoryear{{Meiron}, {Kocsis}  \& {Loeb}}{{Meiron}
  et~al.}{2017}]{meiron_kocsis_2017}
{Meiron} Y.,  {Kocsis} B.,   {Loeb} A.,  2017, \mn@doi [\apj]
  {10.3847/1538-4357/834/2/200}, \href
  {http://adsabs.harvard.edu/abs/2017ApJ...834..200M} {834, 200}

\bibitem[\protect\citeauthoryear{{Miller} \& {Lauburg}}{{Miller} \&
  {Lauburg}}{2009}]{MillerLauburg09}
{Miller} M.~C.,  {Lauburg} V.~M.,  2009, \mn@doi [ApJ]
  {10.1088/0004-637X/692/1/917}, \href
  {http://adsabs.harvard.edu/abs/2009ApJ...692..917M} {692, 917}

\bibitem[\protect\citeauthoryear{{Misner}, {Thorne}  \& {Wheeler}}{{Misner}
  et~al.}{1973}]{1973grav.book.....M}
{Misner} C.~W.,  {Thorne} K.~S.,   {Wheeler} J.~A.,  1973, {Gravitation}

\bibitem[\protect\citeauthoryear{{O'Leary}, {Kocsis}  \& {Loeb}}{{O'Leary}
  et~al.}{2009}]{oleary_kocsis_2009}
{O'Leary} R.~M.,  {Kocsis} B.,   {Loeb} A.,  2009, \mn@doi [\mnras]
  {10.1111/j.1365-2966.2009.14653.x}, \href
  {https://ui.adsabs.harvard.edu/abs/2009MNRAS.395.2127O} {395, 2127}

\bibitem[\protect\citeauthoryear{{Ohanian}}{{Ohanian}}{1973}]{ohanian73}
{Ohanian} H.~C.,  1973, \mn@doi [\prd] {10.1103/PhysRevD.8.2734}, \href
  {https://ui.adsabs.harvard.edu/abs/1973PhRvD...8.2734O} {8, 2734}

\bibitem[\protect\citeauthoryear{{Peng} \& {Chen}}{{Peng} \&
  {Chen}}{2021}]{peng21}
{Peng} P.,  {Chen} X.,  2021, \mn@doi [\mnras] {10.1093/mnras/stab1419}, \href
  {https://ui.adsabs.harvard.edu/abs/2021MNRAS.505.1324P} {505, 1324}

\bibitem[\protect\citeauthoryear{{Pineault} \& {Roeder}}{{Pineault} \&
  {Roeder}}{1977}]{1977ApJ...212..541P}
{Pineault} S.,  {Roeder} R.~C.,  1977, \mn@doi [\apj] {10.1086/155073}, \href
  {https://ui.adsabs.harvard.edu/abs/1977ApJ...212..541P} {212, 541}

\bibitem[\protect\citeauthoryear{{Polnarev}}{{Polnarev}}{1972}]{polnarev72}
{Polnarev} A.~G.,  1972, \mn@doi [Astrophysics] {10.1007/BF01011365}, \href
  {https://ui.adsabs.harvard.edu/abs/1972Ap......8..273P} {8, 273}

\bibitem[\protect\citeauthoryear{{Punturo} et~al.,}{{Punturo}
  et~al.}{2010}]{2010CQGra..27s4002P}
{Punturo} M.,  et~al., 2010, \mn@doi [Classical and Quantum Gravity]
  {10.1088/0264-9381/27/19/194002}, \href
  {https://ui.adsabs.harvard.edu/abs/2010CQGra..27s4002P} {27, 194002}

\bibitem[\protect\citeauthoryear{{Romero-Shaw}, {Lasky}, {Thrane}  \&
  {Calder{\'o}n Bustillo}}{{Romero-Shaw} et~al.}{2020}]{2020ApJ...903L...5R}
{Romero-Shaw} I.,  {Lasky} P.~D.,  {Thrane} E.,   {Calder{\'o}n Bustillo} J.,
  2020, \mn@doi [\apjl] {10.3847/2041-8213/abbe26}, \href
  {https://ui.adsabs.harvard.edu/abs/2020ApJ...903L...5R} {903, L5}

\bibitem[\protect\citeauthoryear{{Rosquist}, {Bylund}  \&
  {Samuelsson}}{{Rosquist} et~al.}{2009}]{2009IJMPD..18..429R}
{Rosquist} K.,  {Bylund} T.,   {Samuelsson} L.,  2009, \mn@doi [International
  Journal of Modern Physics D] {10.1142/S0218271809014546}, \href
  {https://ui.adsabs.harvard.edu/abs/2009IJMPD..18..429R} {18, 429}

\bibitem[\protect\citeauthoryear{{Samsing} et~al.,}{{Samsing}
  et~al.}{2022}]{samsing22}
{Samsing} J.,  et~al., 2022, \mn@doi [\nat] {10.1038/s41586-021-04333-1}, \href
  {https://ui.adsabs.harvard.edu/abs/2022Natur.603..237S} {603, 237}

\bibitem[\protect\citeauthoryear{{Sathyaprakash} \& {Schutz}}{{Sathyaprakash}
  \& {Schutz}}{2009}]{sathyaprakash_schutz_2009}
{Sathyaprakash} B.~S.,  {Schutz} B.~F.,  2009, \mn@doi [Living Reviews in
  Relativity] {10.12942/lrr-2009-2}, \href
  {http://adsabs.harvard.edu/abs/2009LRR....12....2S} {12, 2}

\bibitem[\protect\citeauthoryear{{Schneider}, {Ehlers}  \& {Falco}}{{Schneider}
  et~al.}{1992}]{1992grle.book.....S}
{Schneider} P.,  {Ehlers} J.,   {Falco} E.~E.,  1992, {Gravitational Lenses},
  \mn@doi{10.1007/978-3-662-03758-4.
}

\bibitem[\protect\citeauthoryear{{Schutz}}{{Schutz}}{1986}]{schutz_1986}
{Schutz} B.~F.,  1986, \mn@doi [\nat] {10.1038/323310a0}, \href
  {http://adsabs.harvard.edu/abs/1986Natur.323..310S} {323, 310}

\bibitem[\protect\citeauthoryear{{Stone}, {Metzger}  \& {Haiman}}{{Stone}
  et~al.}{2017}]{stone_metzger_2017}
{Stone} N.~C.,  {Metzger} B.~D.,   {Haiman} Z.,  2017, \mn@doi [\mnras]
  {10.1093/mnras/stw2260}, \href
  {http://adsabs.harvard.edu/abs/2017MNRAS.464..946S} {464, 946}

\bibitem[\protect\citeauthoryear{{Tagawa}, {Haiman}  \& {Kocsis}}{{Tagawa}
  et~al.}{2019}]{tagawa_haiman_2019}
{Tagawa} H.,  {Haiman} Z.,   {Kocsis} B.,  2019, arXiv e-prints, \href
  {https://ui.adsabs.harvard.edu/abs/2019arXiv191208218T} {p. arXiv:1912.08218}

\bibitem[\protect\citeauthoryear{{The LIGO Scientific Collaboration} \& {The
  Virgo Collaboration}}{{The LIGO Scientific Collaboration} \& {The Virgo
  Collaboration}}{2019}]{2019ApJ...882L..24A}
{The LIGO Scientific Collaboration} {The Virgo Collaboration} 2019, \mn@doi
  [\apjl] {10.3847/2041-8213/ab3800}, \href
  {https://ui.adsabs.harvard.edu/abs/2019ApJ...882L..24A} {882, L24}

\bibitem[\protect\citeauthoryear{{The LIGO Scientific Collaboration} \& {The
  Virgo Collaboration}}{{The LIGO Scientific Collaboration} \& {The Virgo
  Collaboration}}{2021}]{gayathri21}
{The LIGO Scientific Collaboration} {The Virgo Collaboration} 2021, \mn@doi
  [\apjl] {10.3847/2041-8213/ac2cc1}, \href
  {https://ui.adsabs.harvard.edu/abs/2021ApJ...920L..42G} {920, L42}

\bibitem[\protect\citeauthoryear{{The LIGO Scientific Collaboration}
  et~al.,}{{The LIGO Scientific Collaboration}
  et~al.}{2021a}]{2021arXiv210801045T}
{The LIGO Scientific Collaboration} et~al., 2021a, \mn@doi [arXiv e-prints]
  {10.48550/arXiv.2108.01045}, \href
  {https://ui.adsabs.harvard.edu/abs/2021arXiv210801045T} {p. arXiv:2108.01045}

\bibitem[\protect\citeauthoryear{{The LIGO Scientific Collaboration}, {the
  Virgo Collaboration}  \& {the KAGRA Collaboration}}{{The LIGO Scientific
  Collaboration} et~al.}{2021b}]{GWTC3}
{The LIGO Scientific Collaboration} {the Virgo Collaboration}  {the KAGRA
  Collaboration} 2021b, arXiv e-prints, \href
  {https://ui.adsabs.harvard.edu/abs/2021arXiv211103606T} {p. arXiv:2111.03606}

\bibitem[\protect\citeauthoryear{{The LIGO Scientific Collaboration}, {the
  Virgo Collaboration}  \& {the KAGRA Collaboration}}{{The LIGO Scientific
  Collaboration} et~al.}{2021c}]{ligo21rate}
{The LIGO Scientific Collaboration} {the Virgo Collaboration}  {the KAGRA
  Collaboration} 2021c, arXiv e-prints, \href
  {https://ui.adsabs.harvard.edu/abs/2021arXiv211103634T} {p. arXiv:2111.03634}

\bibitem[\protect\citeauthoryear{{Thompson}}{{Thompson}}{2019}]{thompson19}
{Thompson} C.,  2019, \mn@doi [\apj] {10.3847/1538-4357/aafda3}, \href
  {https://ui.adsabs.harvard.edu/abs/2019ApJ...874...48T} {874, 48}

\bibitem[\protect\citeauthoryear{{Torres-Orjuela} \& {Chen}}{{Torres-Orjuela}
  \& {Chen}}{2022}]{torres-orjuela22hubble}
{Torres-Orjuela} A.,  {Chen} X.,  2022, arXiv e-prints, \href
  {https://ui.adsabs.harvard.edu/abs/2022arXiv221009737T} {p. arXiv:2210.09737}

\bibitem[\protect\citeauthoryear{{Vecchio}}{{Vecchio}}{2004}]{2004PhRvD..70d2001V}
{Vecchio} A.,  2004, \mn@doi [\prd] {10.1103/PhysRevD.70.042001}, \href
  {https://ui.adsabs.harvard.edu/abs/2004PhRvD..70d2001V} {70, 042001}

\bibitem[\protect\citeauthoryear{{Vijaykumar}, {Kapadia}  \&
  {Ajith}}{{Vijaykumar} et~al.}{2022}]{vijaykumar22}
{Vijaykumar} A.,  {Kapadia} S.~J.,   {Ajith} P.,  2022, \mn@doi [\mnras]
  {10.1093/mnras/stac1131}, \href
  {https://ui.adsabs.harvard.edu/abs/2022MNRAS.tmp.1080V} {}

\bibitem[\protect\citeauthoryear{{Virbhadra} \& {Ellis}}{{Virbhadra} \&
  {Ellis}}{2000}]{virbhadra00}
{Virbhadra} K.~S.,  {Ellis} G. F.~R.,  2000, \mn@doi [\prd]
  {10.1103/PhysRevD.62.084003}, \href
  {https://ui.adsabs.harvard.edu/abs/2000PhRvD..62h4003V} {62, 084003}

\bibitem[\protect\citeauthoryear{{Weber}}{{Weber}}{1970}]{weber70}
{Weber} J.,  1970, \mn@doi [\prl] {10.1103/PhysRevLett.25.180}, \href
  {https://ui.adsabs.harvard.edu/abs/1970PhRvL..25..180W} {25, 180}

\bibitem[\protect\citeauthoryear{{Yang}, {Bartos}, {Haiman}, {Kocsis},
  {M{\'a}rka}, {Stone}  \& {M{\'a}rka}}{{Yang}
  et~al.}{2019}]{2019ApJ...876..122Y}
{Yang} Y.,  {Bartos} I.,  {Haiman} Z.,  {Kocsis} B.,  {M{\'a}rka} Z.,  {Stone}
  N.~C.,   {M{\'a}rka} S.,  2019, \mn@doi [\apj] {10.3847/1538-4357/ab16e3},
  \href {https://ui.adsabs.harvard.edu/abs/2019ApJ...876..122Y} {876, 122}

\bibitem[\protect\citeauthoryear{{Yu}, {Wang}, {Seymour}  \& {Chen}}{{Yu}
  et~al.}{2021}]{yu21}
{Yu} H.,  {Wang} Y.,  {Seymour} B.,   {Chen} Y.,  2021, \mn@doi [\prd]
  {10.1103/PhysRevD.104.103011}, \href
  {https://ui.adsabs.harvard.edu/abs/2021PhRvD.104j3011Y} {104, 103011}

\bibitem[\protect\citeauthoryear{{Zhang}, {Chen}, {Shao}  \&
  {Inayoshi}}{{Zhang} et~al.}{2021}]{zhang21}
{Zhang} F.,  {Chen} X.,  {Shao} L.,   {Inayoshi} K.,  2021, \mn@doi [\apj]
  {10.3847/1538-4357/ac2c07}, \href
  {https://ui.adsabs.harvard.edu/abs/2021ApJ...923..139Z} {923, 139}

\makeatother
\end{thebibliography}




\bsp	
\label{lastpage}
\end{document}